\documentclass[fleqn,usenatbib]{mnras}

\usepackage[T1]{fontenc}
\usepackage{ae,aecompl}

\usepackage{graphicx}	
\usepackage{epsfig}
\usepackage{amsmath}	
\usepackage{amssymb}	
\usepackage{cases}

\newcommand {\delpar}{{\delta_{\rm p}}}

\newcommand {\kms}{\,{\rm km}\,{\rm s}^{-1}}

\newcommand {\mas}{\,{\rm mas}}

\newcommand {\kpc}{\,{\rm kpc}}

\newcommand {\gl}{{l}}
\newcommand {\gb}{{b}}

\newcommand {\Ug}{{U}_g}
\newcommand {\Vg}{{V}_g}
\newcommand {\Wg}{W}
\newcommand {\meanWg}{{\overline{W}}}
\newcommand {\meanUg}{{\overline{U}_g}}
\newcommand {\meanVg}{{\overline{V}_g}}
\newcommand {\Vc}{V_c}
\newcommand {\Uh}{U_h}
\newcommand {\Vh}{V_h}
\newcommand {\Wh}{W_h}

\newcommand {\Rg}{R_{\rm g}}

\newcommand {\Rsun}{{R_{0}}}
\newcommand {\vphisun}{V_{g, \odot}}

\newcommand {\Usun}{{U_{\!\odot}}}

\newcommand {\Wsun}{{W_{\!\odot}}}
\newcommand {\Vsun}{{V_{\!\odot}}}
\newcommand {\Vgsun}{{V_{\!g\odot}}}

\newcommand {\meanW}{{\overline{W}}}

\newcommand {\Lz}{L_{\rm z}}
\newcommand {\Ez}{E_{\rm z}}

\renewcommand{\deg}{{^{\circ}}}

\newcommand {\rad}{{\rm rad}}

\newcommand{\sigpar}{{\sigma_{\rm p}}}
\newcommand{\delsigpar}{{\delta \sigma_{\rm p}}}

\title[$\Lz-\Ug$ wave pattern]{More than just a wrinkle: A wave-like pattern in $\Ug$ vs. $\Lz$ from Gaia Data}

\author[J. Friske \& R. Sch{\"o}nrich ]{
	Jennifer K. S. Friske,$^{1}$\thanks{E-mail: J.Friske@physik.lmu.de}
	Ralph Sch{\"o}nrich,$^{2,3}$
	\\
	$^{1}$Ludwig-Maximilians-Universit\"at, Fakult\"at f\"ur Physik, Schellingstr. 4, 80799 M{\"u}nchen, Germany\\
	$^{2}$University of Oxford, Rudolf Peierls Centre for Theoretical Physics, Clarendon Laboratories, OX1 3PU Oxford, UK\\
	$^{3}$Mullard Space Science Laboratory, University College London, Holmbury St. Mary, Dorking, Surrey, RH5 6NT, UK
}

\date{Accepted XXX. Received YYY; in original form ZZZ}

\pubyear{2019}

\begin{document}
	\label{firstpage}
	\pagerange{\pageref{firstpage}--\pageref{lastpage}}
	\maketitle
	
	\begin{abstract}
		We present a newly found wave--like pattern in mean Galactocentric radial velocity $\meanUg$ vs. guiding centre radius $\Rg$ or angular momentum $\Lz$ of stars in the RV subsample of Gaia DR2. The short-wave pattern has a wavelength of order $1.2 \kpc$ in $\Rg$ or $285 \kpc \kms$ in $\Lz$. The pattern shows only weak changes with Galactocentric radius $R$ and little change in strength in particular with the vertical energy $\Ez$ of the stars or respectively the distance to the Galactic plane $|z|$. The pattern is to first order symmetric around the plane, i.e. has no significant odd terms in $z$. There is weak phase shift with the pattern moving towards slightly lower $\Lz$ (i.e. trailing) with $|z|$ and $\Ez$. However, we observe a highly significant phase shift in Galactic azimuth $\phi$, which is different for different peaks. The peak around $\Lz \sim 2100 \kpc\kms$ only shows a weak change with $\phi$, while the rest of the pattern shows a clearly detectable shift of $\text{d} \Lz /\text{d} \phi = (200 \pm 22) \kpc\kms / \rad$.  If we consider all peaks to belong to the same pattern, this would suggest a wavenumber $m = 4$. We further find that the wave-like pattern in $\Ug$ appears to be related to the $\meanWg$ vs. $\Lz$ pattern detected in Gaia DR1.
		A comparison of the $\meanUg-\Lz$ wave pattern with changes of $\meanUg$ vs. $R$, which have been previously discussed, suggests that the latter can be understood as just the $\meanUg-\Lz$ pattern washed out by blurring (i.e. orbital excursions around their guiding-centre) of disc stars.
		
	\end{abstract}
	
	\begin{keywords}
		stars: kinematics and dynamics --
		Galaxy: disc --
		Galaxy: kinematics and dynamics --
		Galaxy: solar neighbourhood --
		Galaxy: structure
		
	\end{keywords}
	
	
	
	\section{Introduction} 
	
	From its first data release the Gaia satellite mission \citep{gaiaMission16} has been providing a stunning, unprecedented view on disc structure. In fact, we are only very slowly getting to grips with new features in phase space that were not fully anticipated in theoretical models. Given the great technological advance compared to its predecessor Hipparcos \citep{hipparcos97}, we can expect that Gaia's data will reveal many surprising diagnostics of the disc structure and kinematics that are hidden in plain sight. In this paper, we will discuss a newly found dependence of the mean radial velocity $\meanUg$ on angular momentum $\Lz$, which exhibits a strong wave-like pattern.
	
	One example of beautiful surprises by Gaia has been the discovery of vertical substructure in the motion of disc stars. While Hipparcos had just delivered very tentative evidence for the Galactic warp \citep{dehnen1998distribution}, molecular gas observations and star counts of bright stars \citep[e.g.][]{drimmel00, Drimmel01, reyle09} have shown that the Sun is quite close to the line of nodes of a large-scale outer disc warp. We should thus expect stars with large angular momenta to have a positive mean vertical velocity $\meanW$ relative to the disc. Instead of just confirming the tentative evidence of the Galactic warp in Hipparcos \citep[][]{schonrich2018warp, Poggio18}, already Gaia DR1 provided more than a quantification of the warp signal. It revealed the presence of a wave-like pattern with an amplitude of $\sim 1 \kms$ of $\meanWg$ vs. $\Lz$ with a scale-length of order $500 \kpc\kms$ or respectively $\sim 2 \kpc$ in guiding centre radius $\Rg$ \citep[][]{schonrich2018warp}. This wave \citep[confirmed later by][on Gaia DR2]{Kawata18} is likely connected to the much larger waves in the outer disc \citep[][]{Xu15, Bergemann18}: the local amplitude should be a lot smaller due to the much larger local surface density compared to the outskirts of the disc. The most likely culprit is also known: the previous impact of the Sagittarius dwarf galaxy and its then a lot more massive dark matter halo, causing a wake in the halo \citep[][]{Weinberg95} and then the corrugated warp/wave pattern in the Galactic disc \citep[][]{donghia16, Laporte18}.
	
	Related to this vertical wave-structure, Gaia DR2 also brought the discovery of the phase-space spiral \citep{Antoja2018} -- a prominent spiral shape in the $W$-$z$ plane (altitude $z$ above or below the disc plane), which is particularly visible if one colours each bin by its mean (Galactocentric) azimuthal or radial velocity, $\meanVg$ or $\meanUg$. This spiral has been extensively scrutinised \citep{Galah2019} and readily explained with an impact of a companion galaxy and its resulting vertical impulse on the disc, very likely stemming from the last pericentric passage of the Sagittarius Dwarf Galaxy 500-800 Myrs ago \citep{binney2018origin, Laporte2019Sag, DWGaiaSpiral2018}.
	
	Here, we will show that there is a very clear wave-like pattern present in $\meanUg$ vs. $\Lz$ throughout the entire extent (spanning a diameter of more than $4 \kpc$) probed by the Gaia RV dataset. This dataset adds line-of-sight velocities to the Gaia astrometry and thus provides full $6$D phase space information for $\sim 7$ million objects.  In section \ref{sec:data}, we describe the data used for our evaluation and how the catalogue used was derived from the Gaia sample. In the following section \ref{sec:wavepattern}, we present the pattern and investigate its behaviour at different positions in the sample data range. We try to quantify the azimuthal phase shift in subsection \ref{sec:wavenumber} by fitting a suitable function to our data and give an estimate for the involved wavenumbers. Section \ref{sec:LzW} then investigates correspondences between the $\meanUg-\Lz$ pattern and the waves in $\meanWg$ vs. $\Lz$, as well as the trends in $\meanUg$ vs. galactocentric radius $R$. Finally, we summarising our findings in \ref{sec:conclusion}.

	\section{Coordinate frame and definitions}\label{sec:coords}
	
	We use a standard right-handed coordinate system $(U,V,W)$ for the velocities, where $U$ is the component of motion towards the Galactic Centre, $V$ in the direction of Galactic rotation, and $W$ quantifies the motion upwards, perpendicular to the plane.  
	We distinguish between Galactocentric velocity components $(\Ug, \Vg, \Wg)$ which are measured in a cylindrical coordinate system in the rest frame of the Galactic Centre and heliocentric velocity components $(\Uh, \Vh, \Wh)$, which are measured in the Cartesian coordinate system relative to the motion of the Sun. I. e. at the position of any star $\Ug$ points towards the Galactic central axis (and for comparison to the other tradition $\Ug = - v_R$ and $\Vg = v_{\phi}$), while $U_h$ points parallel to the line Sun $-$ central axis. Slightly inconsistent in sign and with the handedness of the coordinate system, we define the angular momentum $\Lz = R \Vg$, positive in the direction of disc rotation. We set the solar Galactocentric radius $\Rsun = 8.27 \kpc$, the total azimuthal velocity of the Sun $\vphisun$, and the Local Standard of Rest velocity vector of the Sun $(\Usun, \Vsun, \Wsun) = (11.1, 12.24, 7.25)\kms$. These values have been taken from \cite{schonrich2012} and \cite{schonrich2010} in accordance with \cite{Gillessen09} and \cite{McMillan17}. The values thus imply a local circular speed $\Vc = 238 \kms$, and a solar value $L_{z\odot} = 2067\kpc \kms$. For each star, we make use of Galactocentric cylindrical coordinates, where $R = \sqrt{x^2 + y^2}$ denotes the in-plane Galactocentric radius, and $(x,y,z)$ are the coordinates in the Galactocentric Cartesian frame with the Galactic Centre (GC) at $R = x = y = z = 0$, $x$ in the radial direction outwards, $y$ in the azimuthal direction, and $z$ perpendicular to the Galactic plane. We denote Galactic longitude and latitude with $(\gl, \gb)$. An important quantity in our analysis is the azimuth of a star, i.e. the in-plane angle between the connection lines Sun--GC and GC--star. The azimuth is taken positive in the direction of rotation with $\phi = 0$ at the solar position.
	
	\begin{figure}
		\centering
		\epsfig{file=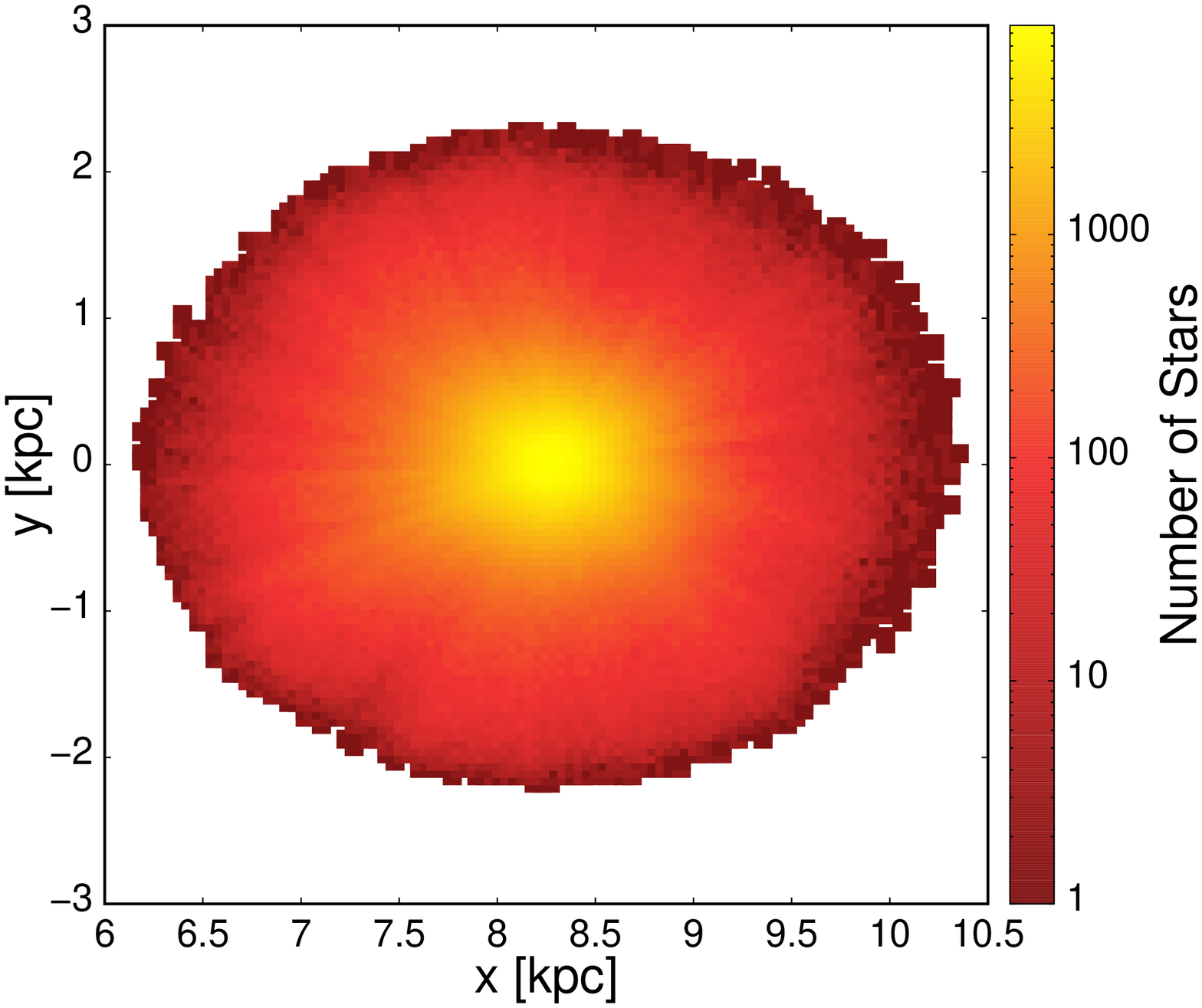,angle=-0,width=\hsize}
		\epsfig{file=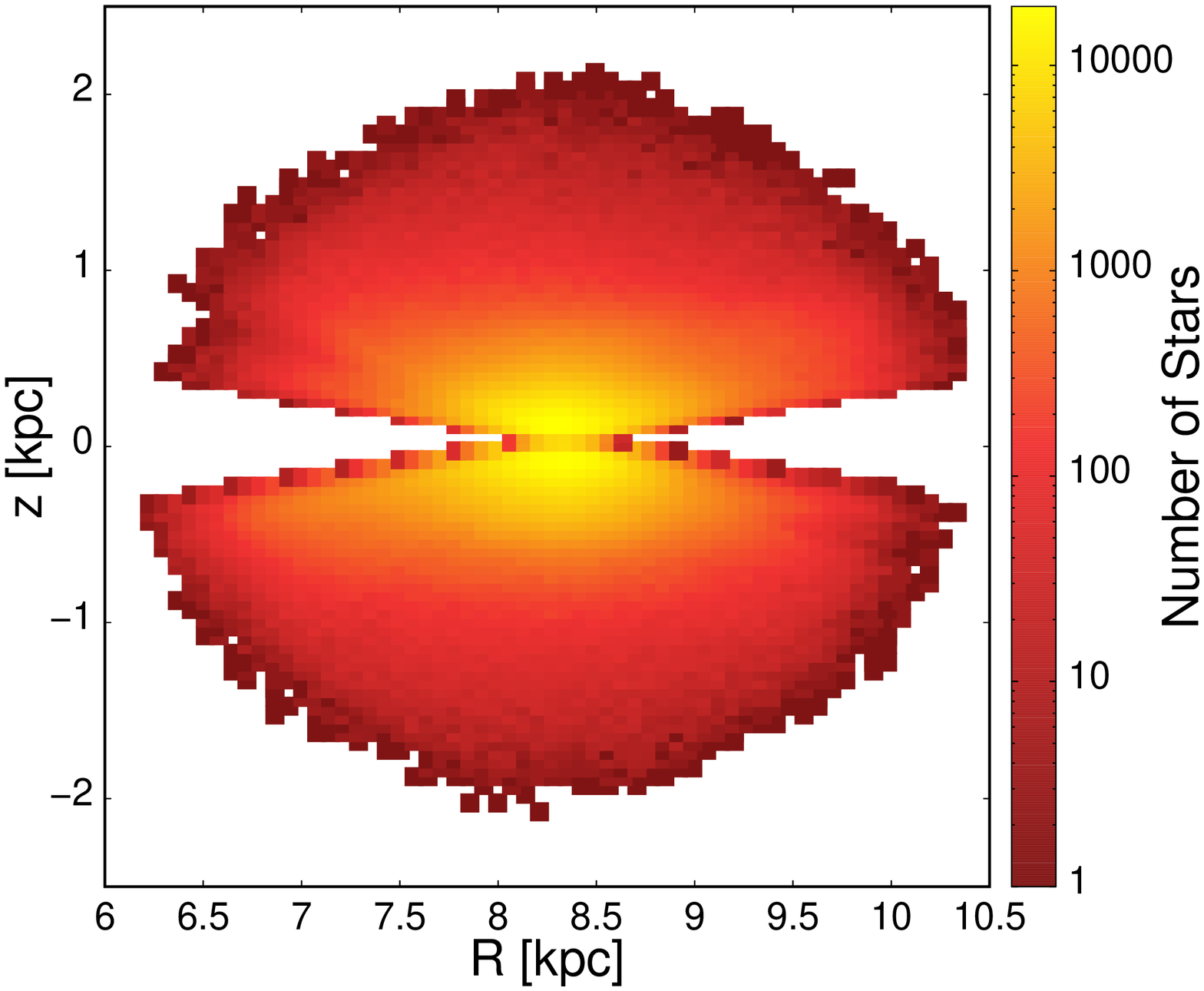,angle=-0,width=\hsize}
		\caption{Geometry of the Gaia Sample we used after introducing the various quality cuts explained in section \ref{sec:data}
		}
		\label{fig:geometry}
	\end{figure}
	
	\section{Data}\label{sec:data}
	
	In this study we make use of the Gaia RV sample \citep[][]{Cropper18, GaiaKatz}, which has been published as a part of Gaia DR2 \citep[][]{GaiaDR2}. Descriptions of the Gaia spacecraft and of its on-board spectrograph can be found in \cite{Prusti16} and \cite{Sartoretti18}. To derive the 6D phase space information, we employ the astrometric information from Gaia DR2, which provides proper motions and parallax measurements (\cite{Arenou18} and \cite{Lindegren18}). For the sake of simplicity, we will stick with the position-velocity phase space for our considerations, however a study of the Gaia DR2 sample in action space can be found in \cite{Trick2019Action}, which also displays some of the ridges and undulations under concern. To provide distances for the sample, we use the distance expectation values from \cite{S19b}, which were derived with the method of \cite{SA17}. Using the expectation values for the distance has the advantage that the expectation values of velocity components should be unbiased. The distances have been extensively tested and calibrated the Gaia parallax offset $\delta p = (0.054 \pm 0.06) \mas$ (Gaia parallaxes are too small), using the method of \cite{SBA}. The implied offset is substantially (a factor $2$) larger than estimated from the Gaia quasar sample in \cite{Lindegren18}. However, the quasar sample is a lot fainter, has a different colour distribution, and has generally zero parallax, factors that advise against a direct extrapolation to the Gaia RV sample. Further, the offset we use is in agreement with tests on particular sources in \cite{Stassun18} and \cite{Zinn18}. We employ the quality cuts suggested in section 8 of \cite{S19b}. We furthermore require a parallax quality ratio $p / \sigpar > 10$ and an additional cut on Galactic latitude of $b > 10°$. The resulting distribution of stellar loci can be found in Fig.~\ref{fig:geometry}.

	\begin{figure}
		\epsfig{file=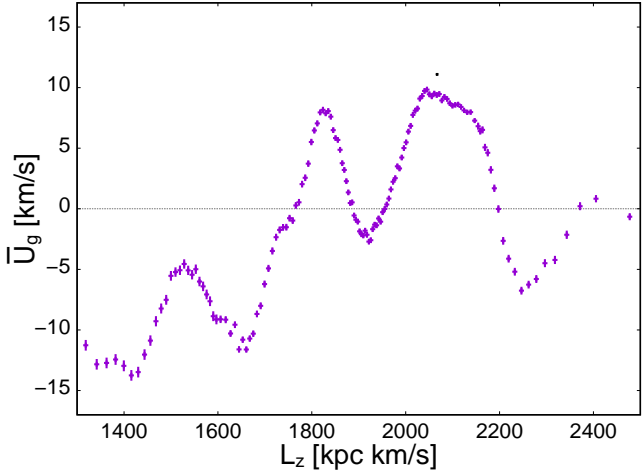,angle=0,width=\hsize}
		\caption{Mean radial velocity component $\meanUg$ of all stars in the sample, when binned by angular momentum $\Lz$. For each data point, $10000$ stars ($20000$ stars for $\Lz$ between $1600\kms$ and $2150\kms$) were taken from the sample sorted in $\Lz$. The error bars depict the Poisson noise in each bin, i.e. $\sigma_{U_g} / \sqrt{N}$, where $N$ is the number of stars in each bin and $\sigma_{U_g}$ its measured dispersion in $\Ug$. }\label{fig:LzVrall}
	\end{figure}

	\begin{figure}
		\epsfig{file=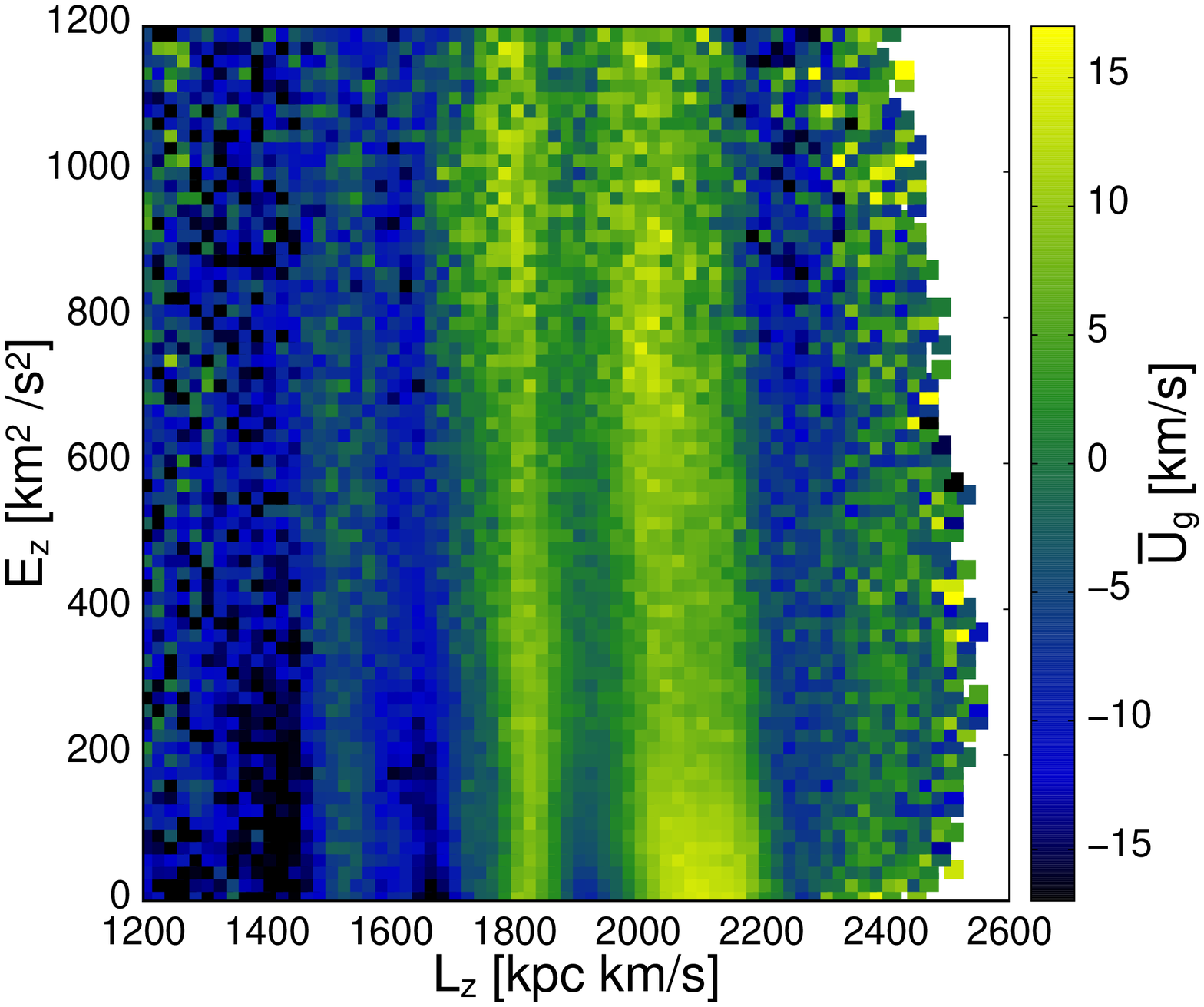,angle=-0,width=\hsize}
		\epsfig{file=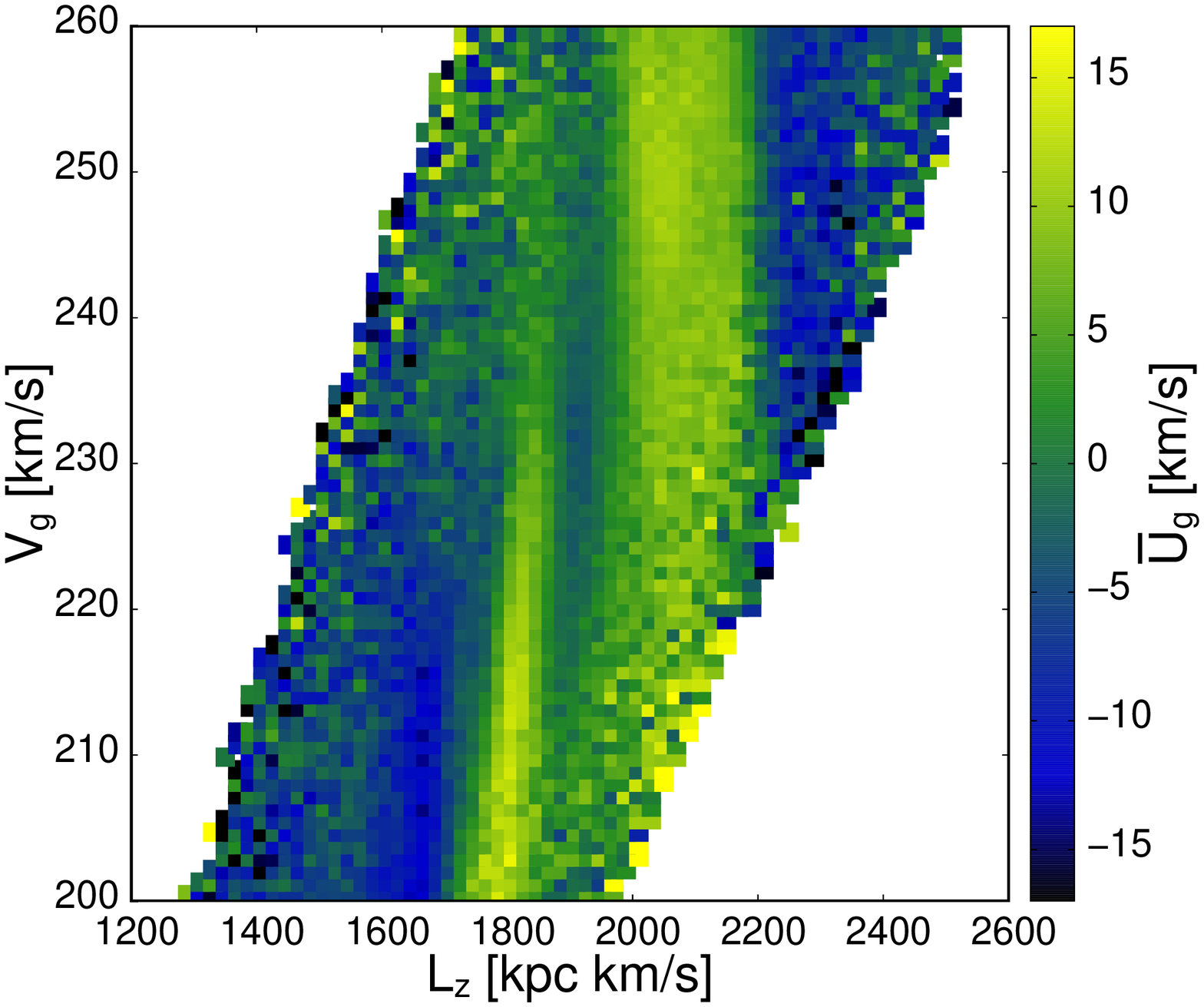,angle=-0,width=\hsize}
		\epsfig{file=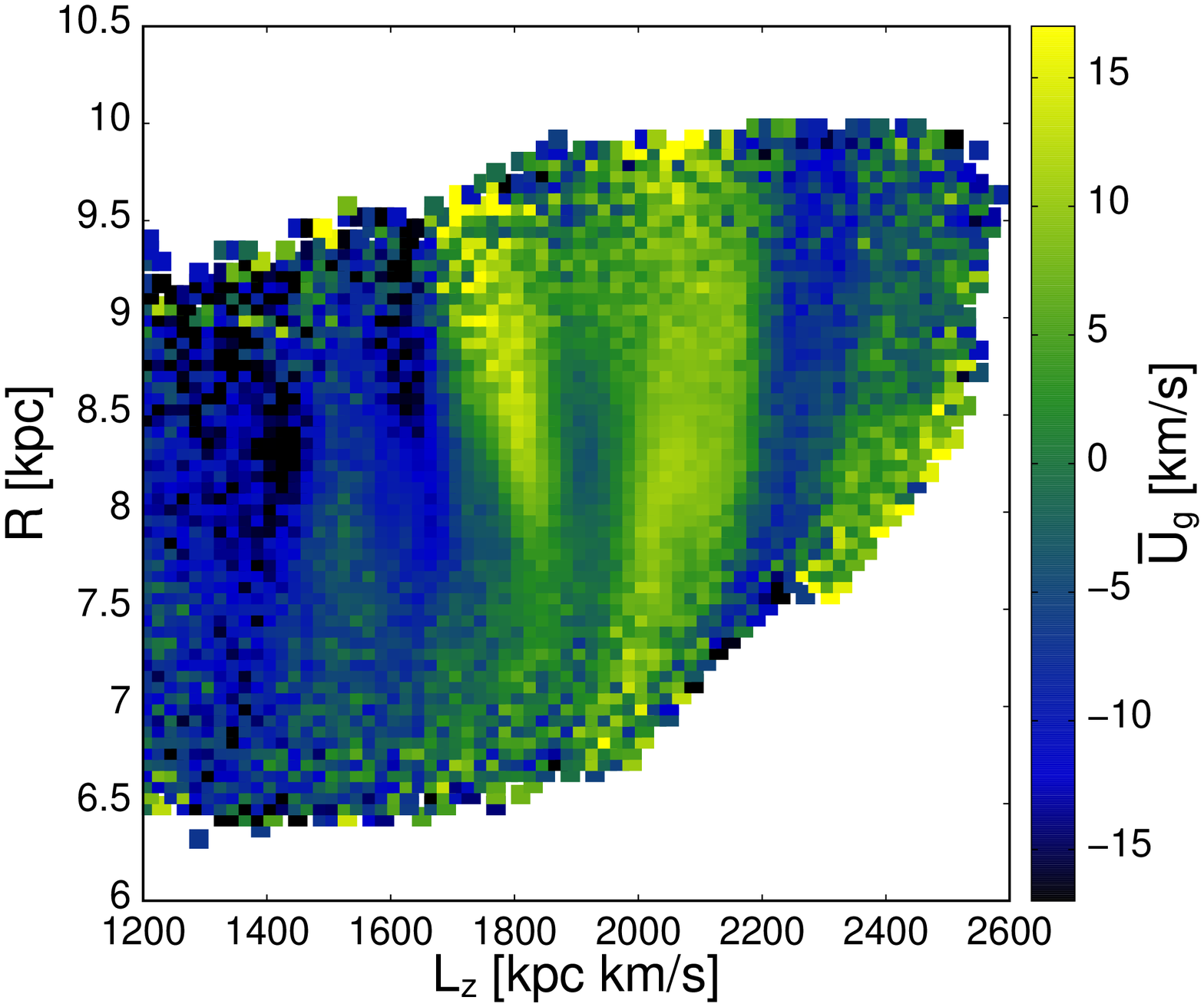,angle=-0,width=\hsize}
		\caption{Mean radial velocity $\meanUg$ marked by colour on the $\Lz$-$\Ez$ plane (top panel), the $\Lz$-$\Vg$ plane (middle) and $\Lz$-$R$ (bottom). For each data bin we required a minimum star number of 10.}\label{fig:LzEz}
	\end{figure}
	
	\section{A new Wave pattern in Gaia DR2} \label{sec:wavepattern}
	
	\subsection{The pattern of radial motion vs. angular momentum} \label{sec:patternradialvsangular}
	
	We can now turn to the topic of this paper: a wave-like structure in the radial velocity component. 
    This wave pattern has occasionally been found in studies of the phase space properties of Gaia DR2, e.g. in Fig. 1 of \cite{Fragkoudi2019} and the lower panel of Fig. 13 in \cite{Laporte2019Sag}. However, it has mostly been pictured in the $R-V$-plane, which obstructs that the most natural variable for its description is the angular momentum $\Lz$. Studying the pattern as a function of $\Lz$ allows us to examine it very closely and even pin down a phase shift throughout our sample.
    
	Fig.~\ref{fig:LzVrall} shows the mean $\Ug$ velocity component in bins of angular momentum. To generate this plot, we ordered the sample in the angular momentum $\Lz = R\Vg$ of each star. For the well-populated region with $\Lz$ between $1600\kms$ and $2150\kms$ we then sliced the sample into bins of $20000$ stars each, while we use bins of $10000$ stars outside the core region in $\Lz$. All data points are formally independent. However, in particular outside the core region in $\Lz$, distance uncertainties will blurr stars modestly along the $x$-axis. The magnitude selection in this sample strongly favours nearby stars, where Gaia parallaxes are very precise. The relative distance uncertainty distribution is provided in Fig.~\ref{fig:GaiaQuality} of the Appendix, and peaks strongly around $2 \%$. Systematic uncertainties will be of order $1 \%$ in the near field and slightly larger for remote stars. 
	
	It is instantly evident from Fig.~\ref{fig:LzVrall} that $\meanUg$ varies with a wave-like pattern of a wavelength of about $300 \kpc \kms$, and an amplitude of order $4 \kms$ in $\meanUg$. The pattern lies on top of larger scale fluctuations that are of order $10 \kms$, i.e. we see that stars with low angular momentum have a net outwards motion relative to the LSR. 	
	The general trend of $\meanUg$ with $\Rg$ and hence $\Lz$ has been expected from prior models analysing the effects of the bar: Resonances of the bar are not only identified with features like the Hercules-stream \citep[][]{Dehnen00, Perez17, Monari19} (which itself has a strong average radial motion outwards), but the bar is also expected to cause a significantly non-zero $\meanUg$ all around the disc on its main resonances \citep[][]{Muehlbauer03}, most notably the change of sign from negative to positive $\meanUg$ shortly outside the outer Lindblad-resonance due to the quadrupole moment of the bar. Similarly, we expect some imprints of the spiral arms both in $\meanUg$ vs. $\Lz$ and in $\meanUg$ vs. $R$  \citep[e.g.][]{monari2016Spiral}. However, the wave--like nature of the pattern in $\meanUg$ is unexpected, and somewhat surprising. We can make out in total $4$ peaks in $\meanUg$, where the peak near the solar value of $\Lz \sim 2070 \kpc $ is wider and has a different appearance. We also experimented with different sample sizes, but it appears impossible at current stage to resolve more extreme values of $\Lz$. 
	
	The first two questions to ask are of course: How confined is this feature to the plane, and, given that we detect it in $\Lz$ only, how confined is it in Galactocentric radius? This is answered in Fig.~\ref{fig:LzEz}. To create the top panel, we introduced the local vertical energy relative to a star moving with no vertical velocity in the mid-plane:
	\begin{equation}
	\Ez = \frac{1}{2} \Wg^2 + \Phi(R,z) - \Phi(R,0) \text{ ,}
	\end{equation}
	where $\Phi(R,z)$ is the galactic potential according to the mass model of \cite{McMillan17} at the position of the star, $\Phi(R,0)$ is the potential at its projected position in the plane, and $\Wg^2$ is the square of its vertical velocity component. Of course the vertical action would be a better conserved quantity, but $\Ez$ provides a straight-forward ordering in vertical extent along each orbit. We can see that the pattern persists with little change towards larger $\Ez$, even though the maximum vertical energy plotted here corresponds to a vertical velocity of $\Wg \sim 49 \kms$ or respectively orbits extending to $z \sim 1 \kpc$ altitude. In light of this stability, we expect the same result in vertical action. Curiously, the pattern slightly shifts in $\Lz$ with $\Ez$: most prominently, the large peak around $\Lz \sim 2100 \kpc \kms$ shifts to of order $80 \kpc \kms$ lower $\Lz$ for large $\Ez$ values. Similar trends can be guessed for most of the remaining pattern, though the highest $\Lz$ feature hints at a slight rightward drift relative to the rest of the pattern, with the minimum of $\meanUg$ around $\Lz \sim 2250 \kpc\kms$ deepening somewhat towards larger $\Ez$. The left-ward drift can be expected if the peaks are physically connected to orbital resonances, since the orbital frequencies should show a slight drift downwards for vertically more extended orbits.
	
	The middle and bottom panels of Fig.~\ref{fig:LzEz} show the sample in the $\Lz$-$\Vg$ and $\Lz-R$ planes. Of course, both plots are closely related by $\Lz = R\Vg$. Lines of constant radius in the middle panel run from lower--left to upper--right. The strong selection bias of the sample towards the Solar Galactocentric radius implies that we have a darth of stars in the top left and bottom right corners of this plot. There is some dependence of the pattern on $\Vg$. The most likely interpretation is that the wave-pattern has some dependence on the orbital phase, although at current stage it is difficult to disentangle this completely from sample selection effects. The sample selection shows more strongly in $\Lz$ vs $R$ (bottom panel). Again there is a dependence on $R$, mostly in amplitude, but also with a surprising curving of the pattern at $R \sim 7.5 \kpc$. A part of the amplitude dependence can be ascribed to the asymmetry in distance biases.
	
	\begin{figure*}
		\centering
		\epsfig{file=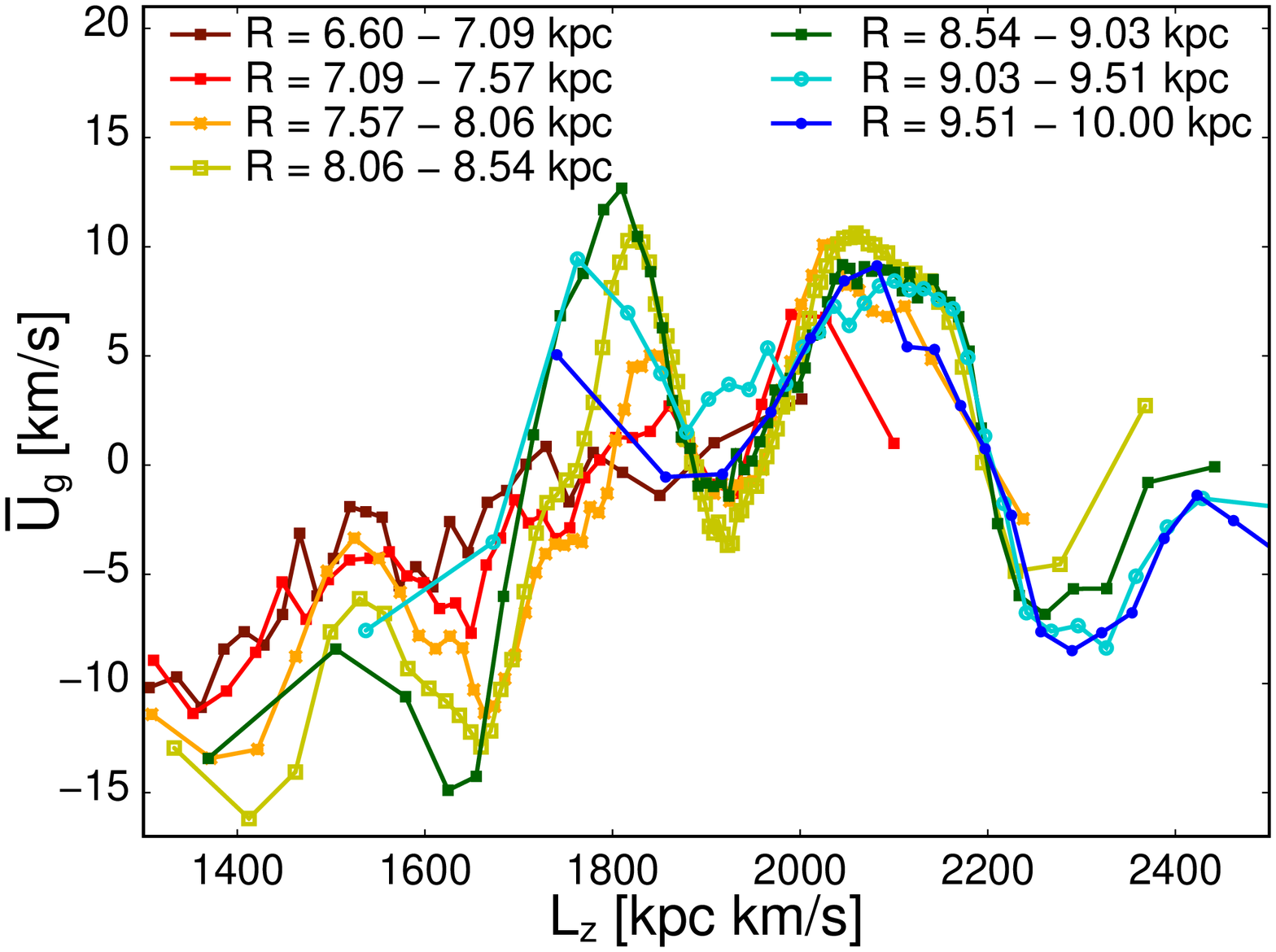,angle=-0,width=0.49\textwidth} 
		\epsfig{file=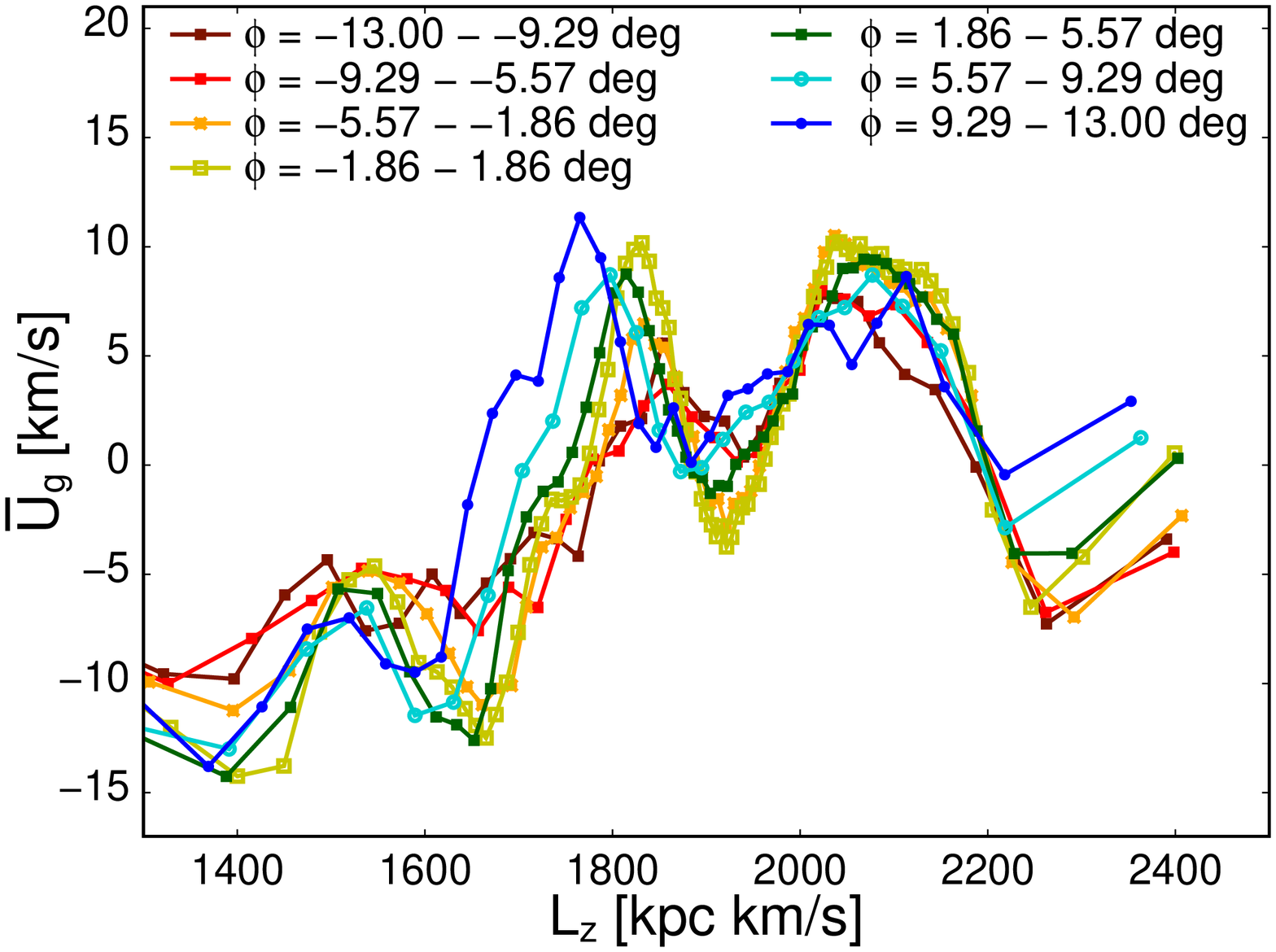,angle=-0,width=0.49\textwidth} 
		\epsfig{file=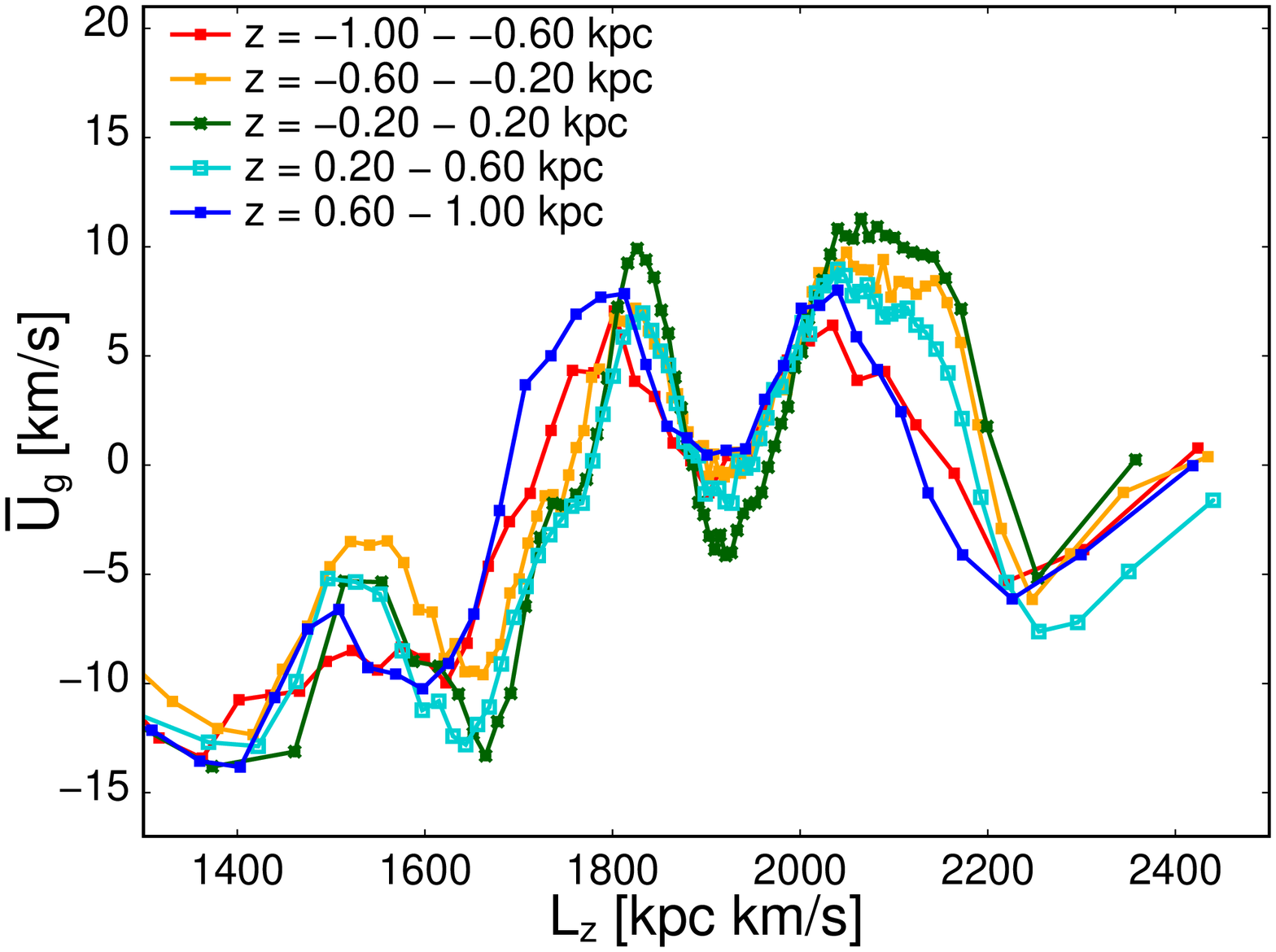,angle=-0,width=0.49\textwidth} 
		\epsfig{file=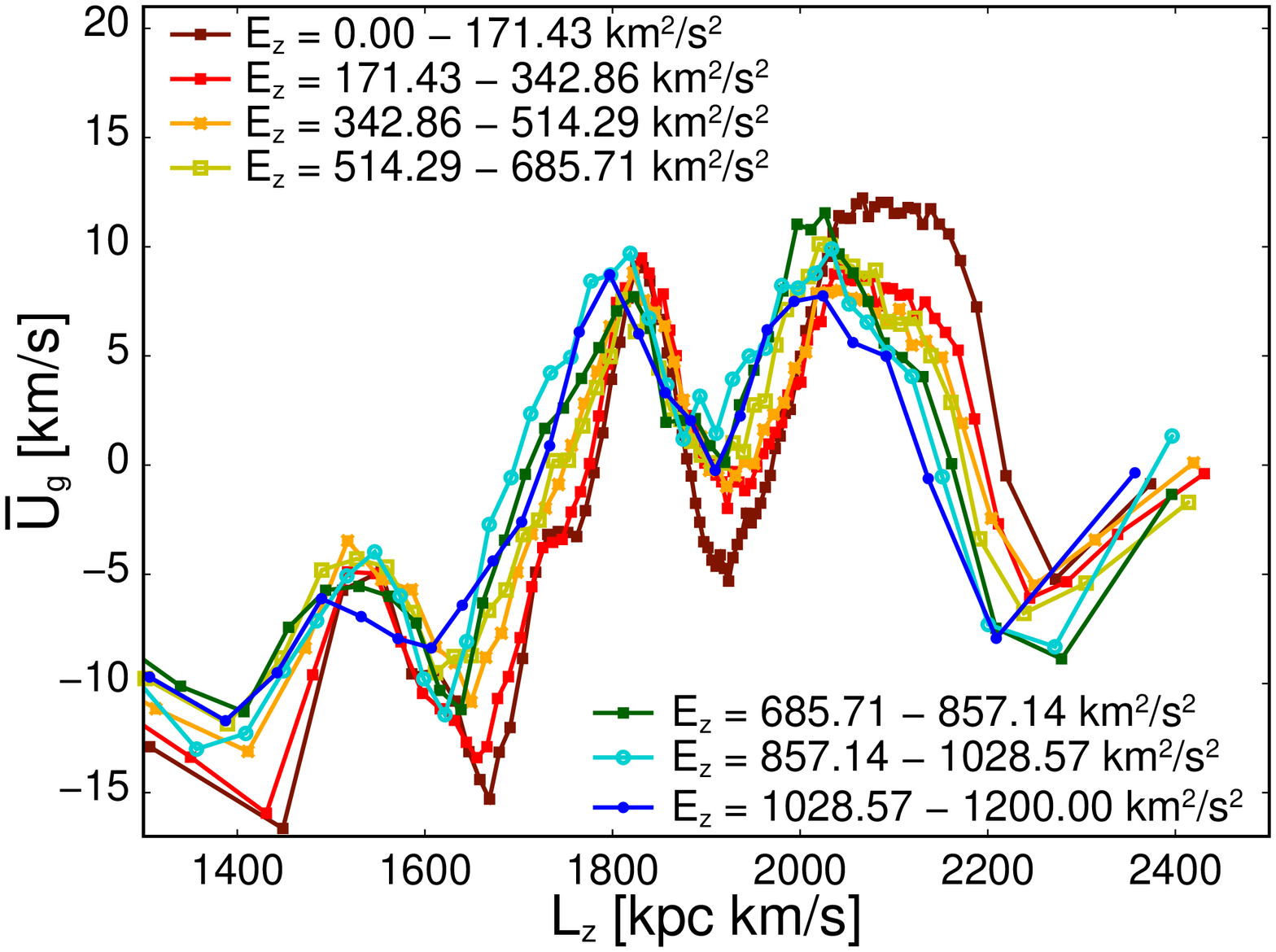,angle=0,width=0.49\hsize}
		\caption{We use the same data as in Fig.~\ref{fig:LzVrall}, but divide the sample into equal width regions in Galactocentric radius $R$ (top left panel), Galactic azimuth $\phi$ (top right), vertical altitude $z$ (bottom left), and vertical energy $\Ez$ (bottom right).
		}
		\label{fig:diffstats}
	\end{figure*}
	
	To get a better grip on the $R$ dependence and other parameters, we repeat the plot of Fig.~\ref{fig:LzVrall} in Fig.~\ref{fig:diffstats} when selecting different ranges in each parameter. The top left panel of Fig.~\ref{fig:diffstats} shows $\meanUg$ vs. $\Lz$ for different bands of $R$. It is clear that for large $R$ we have to lose all signal at low angular momenta, and for small $R$ we have virtually no stars with large angular momentum. As indicated in the previous paragraph, for small $R < 7.5 \kpc$, the wave-pattern appears to get consistently weaker, especially in the low $\Lz$ range. However, this range is plagued by the larger uncertainties for remote stars, which blurr out the peaks in $\meanUg$. 
	
	This effect is also asymmetric in $R$ and in $\Lz$, which we make clear with an example: at large distances of order $2 \kpc$, a realistic Gaia parallax error of $\sigpar \sim 0.05 \mas$ results in a distance uncertainty of $\sim 10 \%$. In direction towards the Galactic Centre, a distance error affects the $\Lz$ measurement twice in the same direction: a star with $(R,\Vg) = (6 \kpc, 200 \kms)$ will, if we underestimate its distance by $10 \%$, be measured at $(R',\Vg') \sim (6.23 \kpc, 205 \kms)$, or respectively will have moved about a quarter wave-length of the wave-pattern in $\Lz$. If we measure a star with high $\Vg > 250 \kms$ in the same place, the $\Vg$ velocity error will change sign and counteract the error in $R$. Analogously, errors for stars observed at large radii (towards the Galactic anticentre) with low $\Vg$ tend to cancel, while they will be additive for stars with large $\Vg > \Vgsun$. This explains the main trends in $\meanVg$ and $R$. We have formally good data for low $\Lz$ and low $\Vg$, but in particular at small $R$, the additive error budget will smear out the pattern in $\Lz$. Similarly, the wave--pattern disappears at large $R$ and small $R$, explaining the loss of signal at extreme values. 
	In short, while our datasets have been optimised for unbiased distances, we still face a loss of signal to random errors.
	
	The bottom left panel of Fig.~\ref{fig:diffstats} shows the wave-like pattern sliced in different bins of $z$. Focusing e.g. on the two most extreme altitude bands at $0.6 < |z|/\kpc < 1$ vs. the rest of the sample, we can see that the pattern shifts symmetrically with increasing altitude. Note that the Galactic latitude cut of $\gb > 10^\deg$ implies that for small $|z|$ our sample is restricted to near solar radius stars. However, the quality difference with random errors to first order only affects the measured amplitude, so the consistent pattern/phase shift with $|z|$ is real. Further, the symmetry in $z$ of the shift argues against systematic distance errors. The left-shift of the pattern can quite straight-forwardly be attributed to the slower orbital frequencies for stars reaching high altitude.  
	
	The phase shift with $|z|$ is consistent with the bottom right panel in Fig.\ref{fig:diffstats}, where we separate the sample in the vertical energy $\Ez$ of the stars. Again, with increasing vertical extent of the orbits, the pattern exhibits a constant shift to lower angular momentum. The shift is clearly detectable through the entire $\Lz$ range. We find again that with increasing $\Ez$ the signal tends to get weaker. Yet, again at least some of this loss of amplitude will be caused by the increasing distance uncertainties.
	
	The top right panel tries to resolve the $\phi$ dependence of the wave-like structure. This is particular interesting, since it can give us clues about the perturbation causing the pattern and the wavenumber of this perturbation. The phase shift with $\phi$ is already visible to the naked eye. In particular, the left prominent peak shifts systematically to the left with increasing $\phi$, i.e. resembles a trailing pattern. Again, measurements at large $|\phi|$ are ridden with distance uncertainties, since all their stars are at the fringes of our quality limits, but this is a promising start for a further investigation.
	Unlike $|z|$, where all peaks shifted consistently, the behaviour in $\phi$ indicates that this is a superposition of at least two causes/modes with different wavenumber. 
	
	\begin{figure}
		\centering
		\epsfig{file=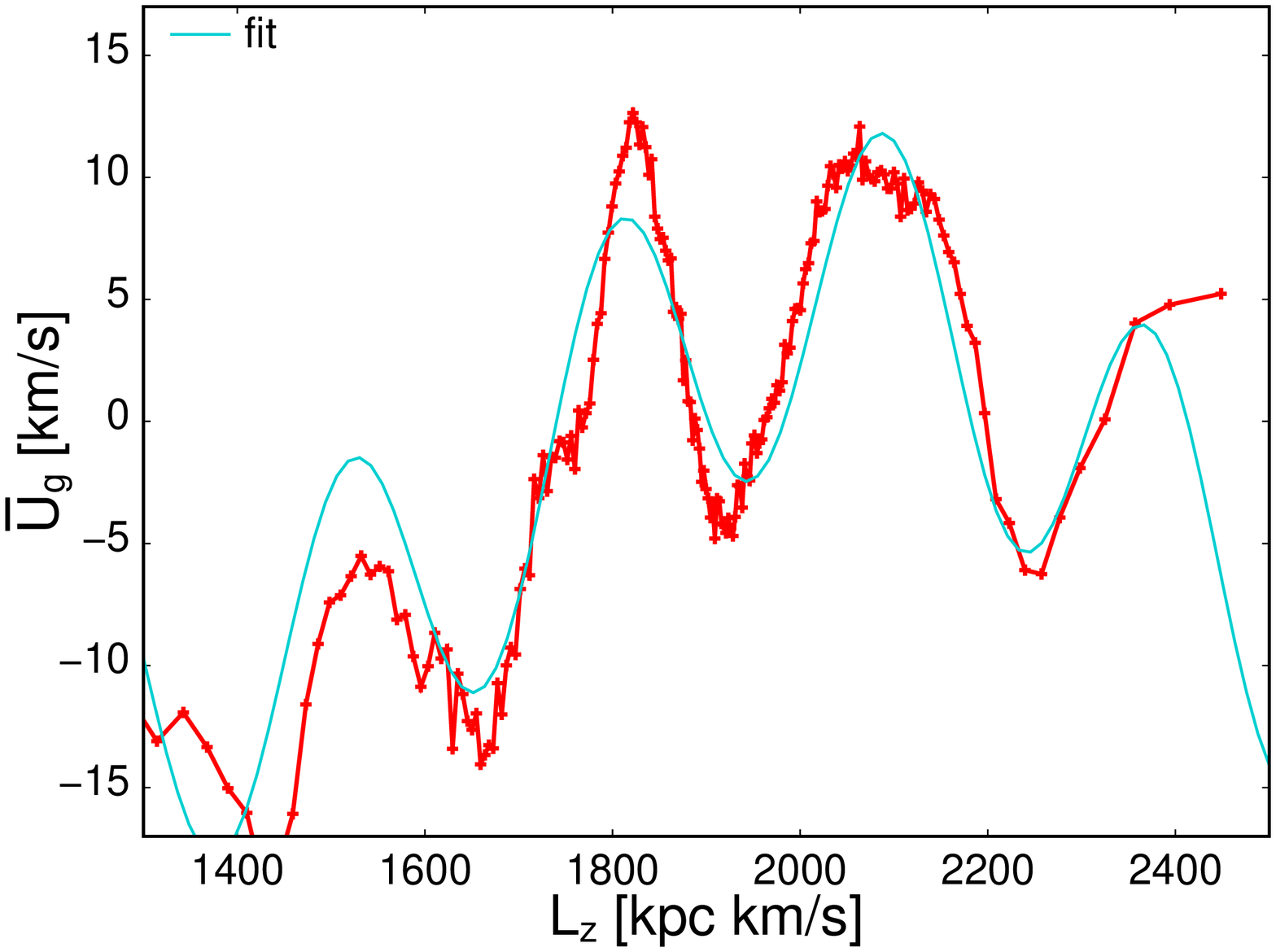,angle=-0,width=\hsize}
		\epsfig{file=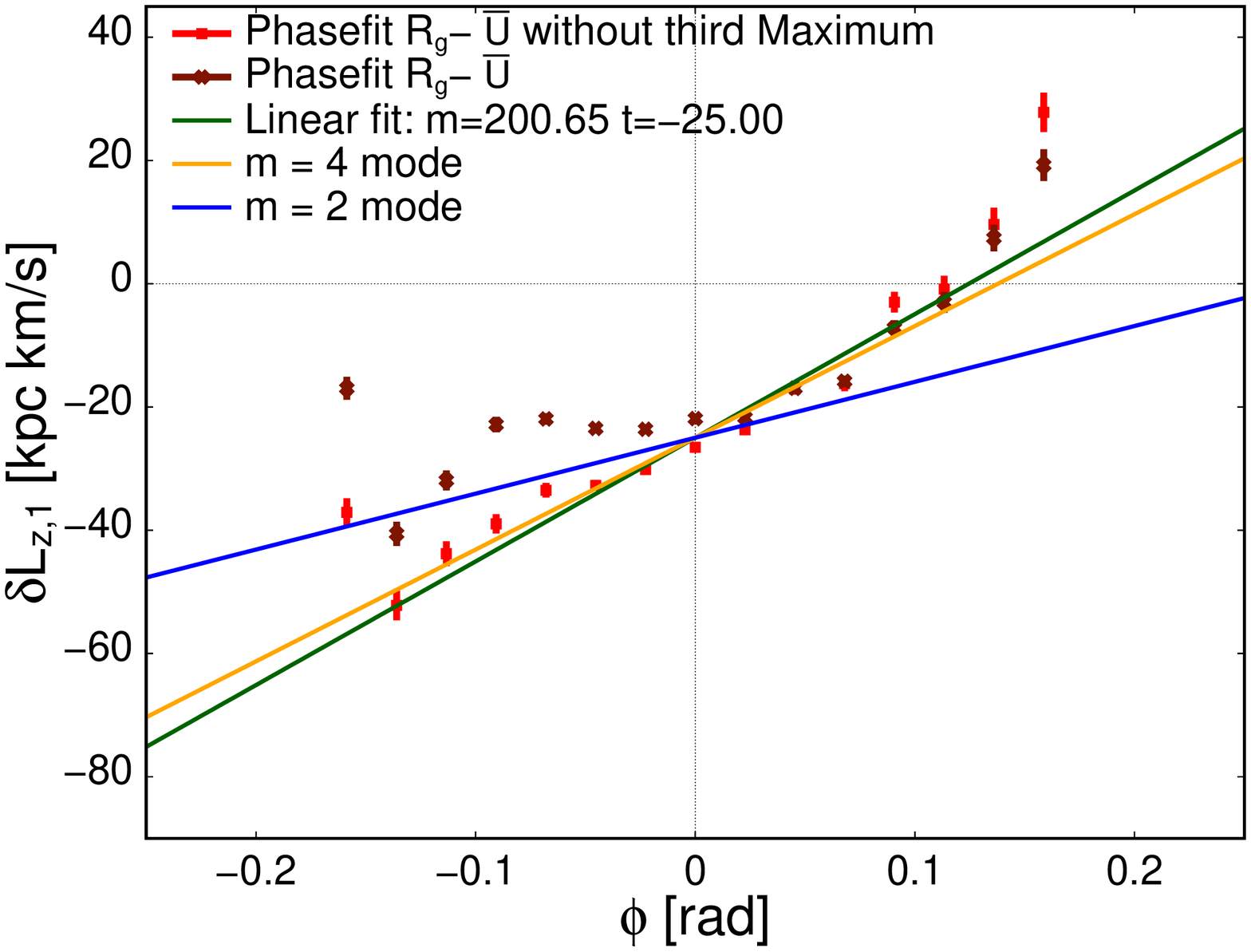,angle=-0,width=\hsize}
		\caption{Top panel: Fit of function \ref{eq:key} to a data sample restricted to a box of radial width of $0.49\kpc$ and angular width of $3.7^\circ$ centred around the sun. 
			Bottom panel: 
			The phase shift $\delta_1$ via eq. \ref{eq:key} with increasing azimuth. 
			We compare two fits, one for a sample including the whole $\Lz$ range and one fit where we exclude the angular momentum region of the third peak between $1950 \kpc \kms$ and $2250 \kpc \kms$.}
		\label{fig:phasefit}
	\end{figure}

	\subsection{Constraining the phase shift}
	\label{sec:wavenumber}
	Just having seen the phase shift of the pattern with Galactic azimuth $\phi$, we attempt to quantify it and to derive the wavenumber $m$. 
	Any fit will be far from perfect, since the changing distance error distribution will increasingly wash out the wave pattern towards larger $|\phi|$. 
	The idea is that if we find a half-way decent fit function for a local, well-constrained part of the sample that also gives reasonable fits for the adjacent regions, the alignment of data and the fit should be enough to pick out the average amount of phase shift, even with imperfect parameters for amplitude and wavelength in $\Lz$. 
	If we then assume that we have made a reasonable choice for the wavelength of the fit function, we can estimate the wavenumber $m$ from the slope of the fit in $\delta L_{z}$ vs. $\phi$ by dividing it by the wavelength. 
	
	To find the best-guess fit function, we picked a subsample restricted to $8.06-8.55 \kpc$ in Galactocentric radius and $-1.86$ to $1.86 ^\circ$ in azimuth, which is depicted in the top panel of Fig.~\ref{fig:phasefit}.
	As can be seen from the blue line in the plot, the pattern is well-fit by a superposition of a fast and a slow wave:
	
	\begin{equation}
	\label{eq:key}
	\meanUg(L_z) = \sum_{i=1}^2{ A_i \sin\left((L_z + \delta_{ L_{z,i}}) \cdot \frac {2 \pi}{\lambda_i} \right) + t {\rm .}}
	\end{equation}
	
	We held constant our amplitudes $A_1= 7.0 \kms$, $A_2 = 8.0 \kms$ and wavelengths $\lambda_1 = 285 \kpc \kms$, $\lambda_2=1350 \kpc \kms$, $\delta_{L_{z,2}} = 1010 \kpc \kms$ and $t = -2.877$ km s$^{-1}$ and varied the phase $\delta_{L_{z,1}}$ for azimuthal slices of width $1.68^\circ$, fitting to the full sample in each range of $\phi$.
    Fixing the $\delta_{L_{z,2}}$ of the slowly varying wave is a matter of fitting hygiene. The pattern is not a perfect sinusoid, and biases and uncertainties wash it out near the rim of the sample. Leaving all parameters free would lead to wild jumps in parameters, with the fitted function aligning with other features than the main ridges/troughs identified here. Since the main objective of this fit is to constrain the phaseshift of the specific extrema that seem to be guided by the same wavenumber, and lacking for now a comprehensive physical model that reproduces the precise shape of the pattern, we allow ourselves the simplification of fitting only the phase of the short wavelength wave.
	
	The lower panel of Fig.~\ref{fig:phasefit} shows the fitted $\delta_{L_{z,1}}$ over the average azimuth in their bin and assumed a linear fit to derive the average phase shift in $\kpc \kms \rad^{-1}$ as slope of that fit. We found that taking the whole sample into account resulted in a slope of $ (97 \pm 32) \kpc \kms \rad ^{-1}$. 
	
	\begin{figure}
		\centering
		\epsfig{file=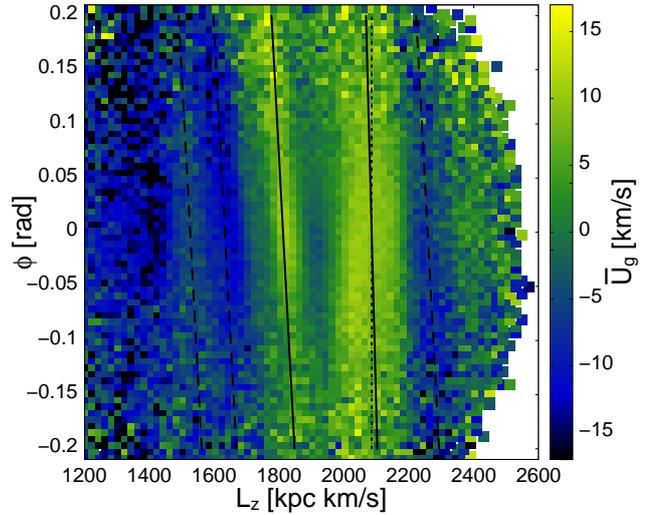,angle=-0,width=\hsize}
		\caption{The mean $\Ug$ velocity component in bins of $\phi$ and $\Lz$. We can see the systematic phase-shift of the single peaks and troughs. It becomes quite clear that the most prominent peak appears to have a lower wavenumber $m$  than the rest of the pattern. 
			To guide the eye, we have entered lines for phase shifts in $\Lz$ with $m = 4$ to the peak at $\Lz(\phi=0) = 1820 \kpc \kms$ and $m=2$ as well as a vertical dot-dashed line for the wide peak at $\Lz(\phi=0) = 2100 \kpc \kms$. 
		In addition the main maxima, we have also introduced lines corresponding to wavenumbers of $m=4$ for three more extrema (dashed lines).
		}
		\label{fig:phaseplane}
	\end{figure}
	
	To help a further analysis of the phase shift, Fig.~\ref{fig:phaseplane} shows the complete azimuthal plane of our sample. It is clear that the two prominent maxima behave differently with the left one shifting more rapidly to large $\Lz$ for small azimuth. This is also the main problem when fitting, as the maxima nearly coincide at small $\phi$. Together with the increased distance uncertainties, fits at large $|\phi|$ become unreliable. The peak near $\Lz \sim 2100 \kpc \kms$ shows again a clearly different behaviour from the rest of the pattern, with a significantly lower phase shift in $\phi$. We therefore re-fitted eq.~\ref{eq:key}, while excluding stars with $1950 < \Lz / (\kpc \kms) < 2250$. The resulting phase shifts are shown with bright red error bars in the bottom panel of Fig.~\ref{fig:phasefit}). The best-fit linear slope for these (shown with the green line) is $d \Lz / d \phi = (202 \pm 22) \kpc \kms \rad^{-1}$. Again, the different results for the phase shift after excluding the broad peak in $\meanUg$ from the sample indicates that two causes with different wavenumbers are responsible for the pattern. If we believe the wavelength of the fit function, the phase shift translates to a wavenumber $m = 4.4 \pm 0.5$, suggesting an $m = 4$ perturbation.
	
	We tried to make this more evident by introducing lines in Fig.~\ref{fig:phaseplane} depicting different linear phase shifts $d \Lz / d \phi$ at the positions of the maxima. 
	We can see that indeed the maximum near $\Lz \sim 1850 \kpc \kms$ is well approximated by a phase shift of $180 \kpc \kms \rad^{-1}$ (consistent with a  $m=4$--mode). The broad peak around $\Lz \sim 2100 \kpc \kms$ is too wide to allow for a firm statement. It is incompatible with the large $d \Lz / d \phi$, but consistent with any scenario between a moderate $d \Lz / d \phi = 90 \kpc \kms \rad^{-1}$ (solid line, consistent with a $m=2$--mode) or even a constant phase (dot-dashed line).
    For three other, less prominent extrema, we also introduced dashed lines corresponding to a wavenumber of $m=4$. Whereas the right minimum is not incommensurable with this phaseshift, the proximity to the wide maximum to its left renders it hard to make a definite claim for its phaseshift. The two extrema at low $\Lz$ however are again well fitted by the $m =4$ mode. 
	
	We conclude that the observed pattern is consistent with a superposition of two trailing patterns in Galactic azimuth. 
	
	\begin{figure*}
		\centering
		\epsfig{file=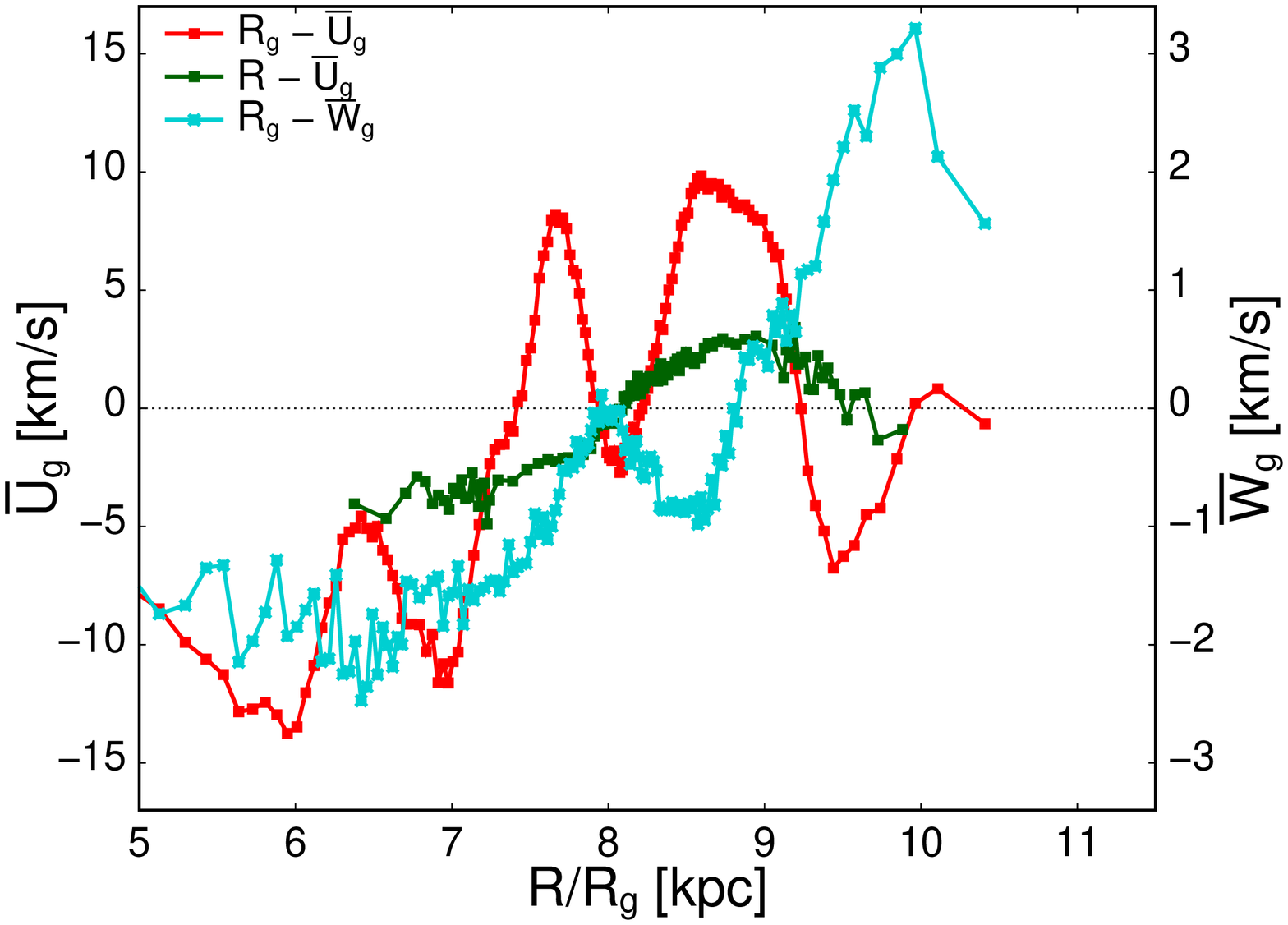,angle=-0,width=0.49\hsize}
		\epsfig{file=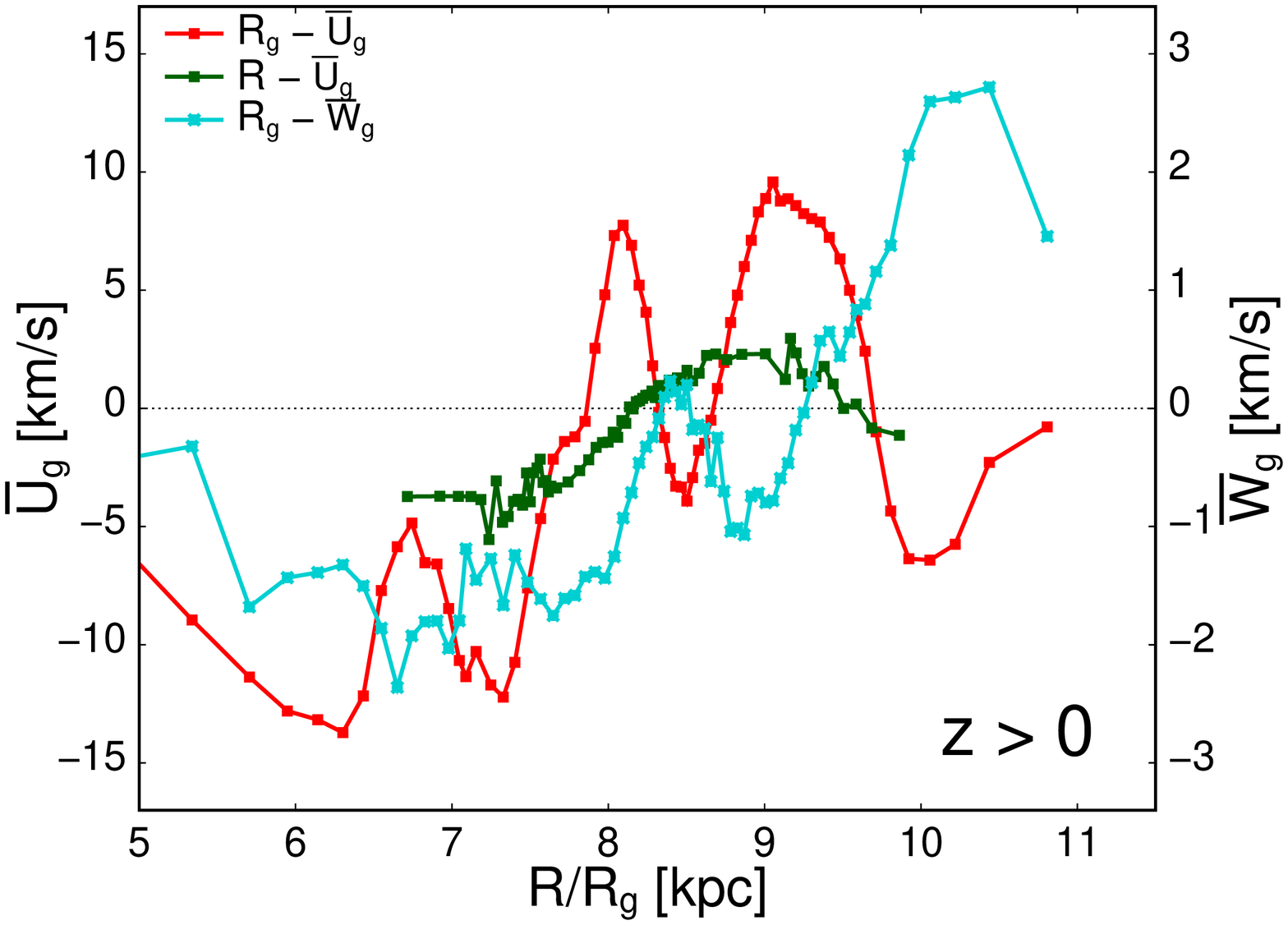,angle=-0,width=0.49\hsize}
		\epsfig{file=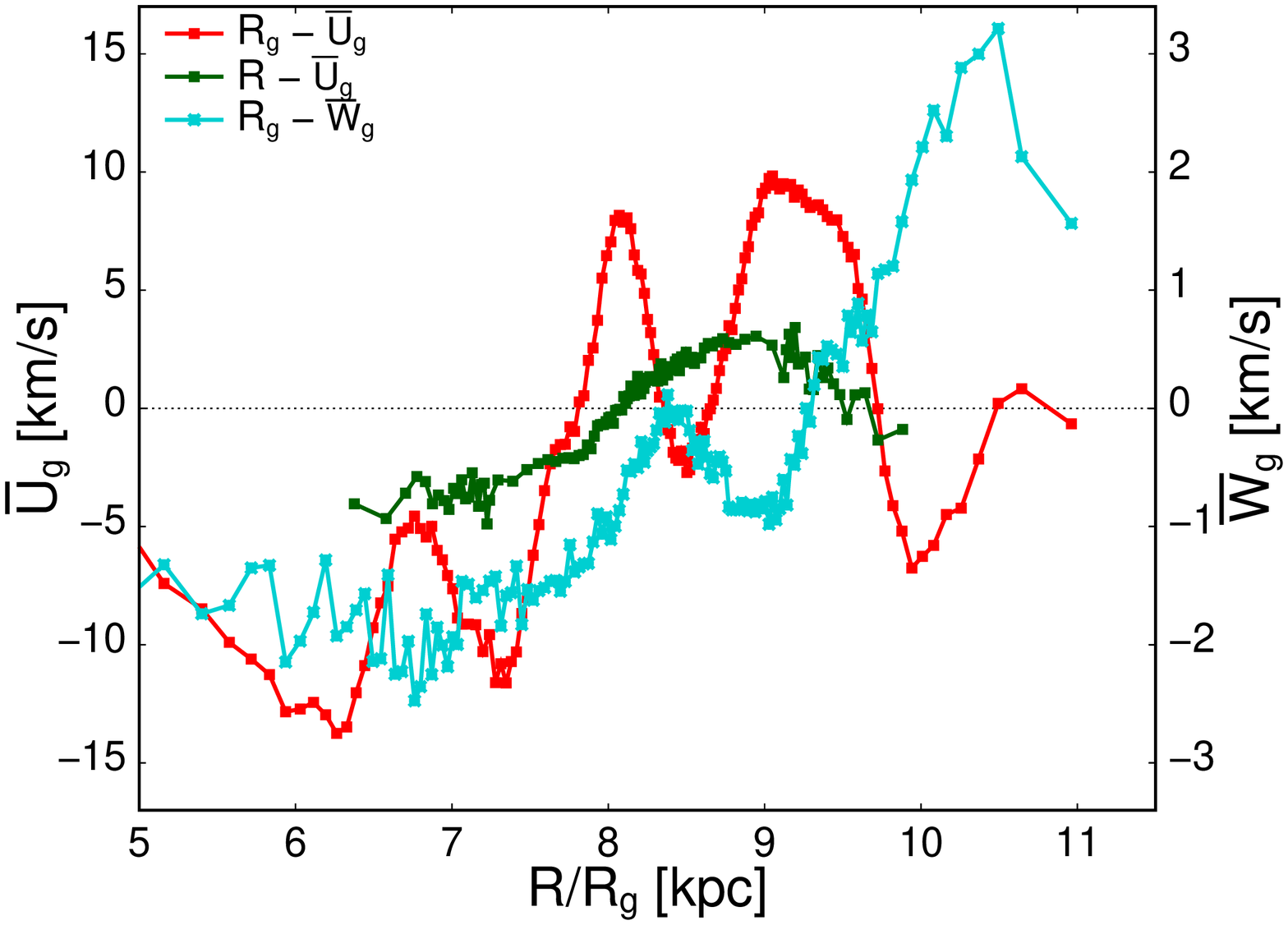,angle=-0,width=0.49\hsize}
		\epsfig{file=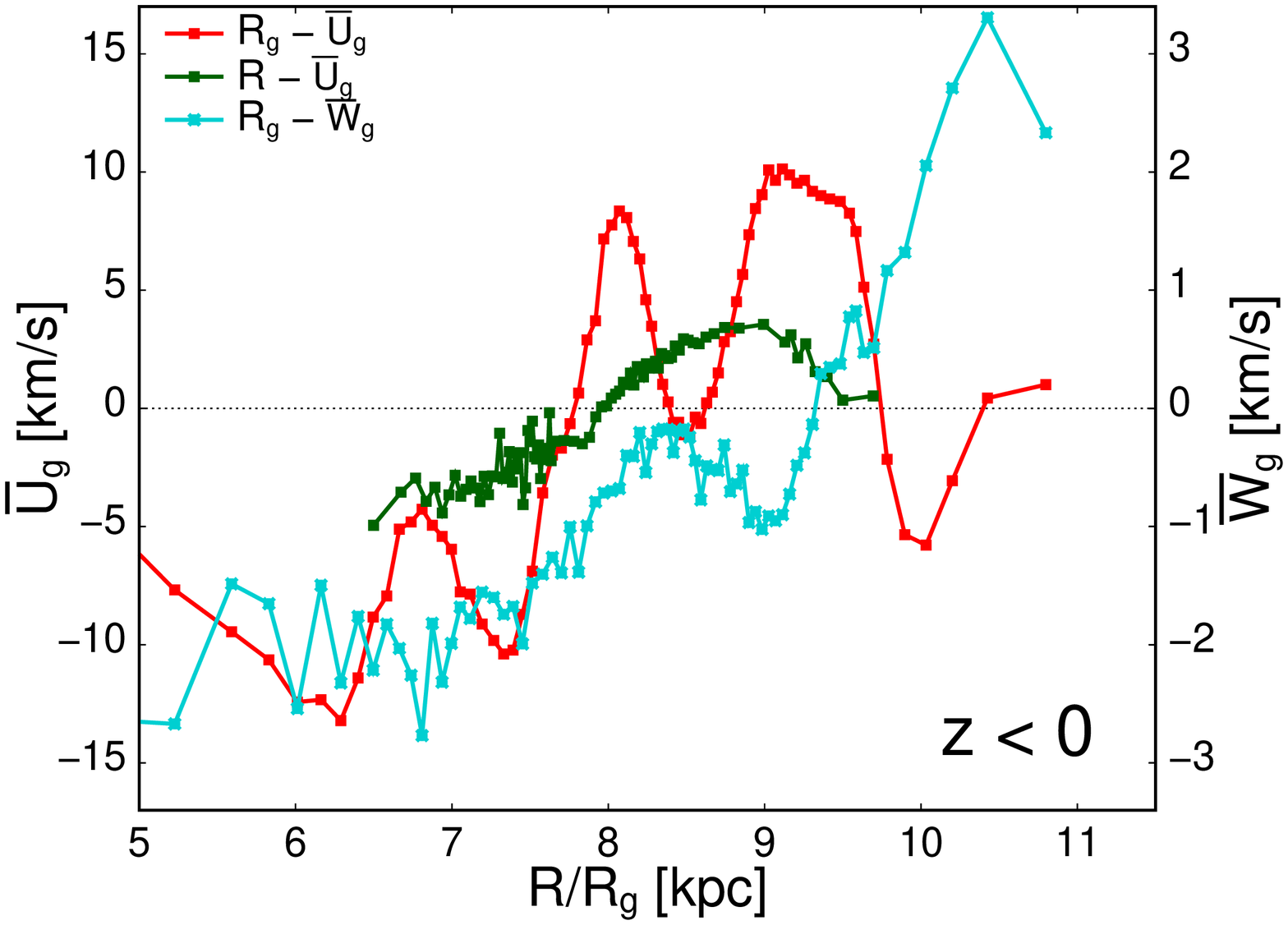,angle=-0,width=0.49\hsize}
		\caption{Comparison of the signal in $R-\meanUg$ and the one in $R_g-\meanUg$ and $R_g-\meanWg$. To highlight the alignments of the extrema and make the weaker $R_g-\meanWg$ signal better visible we stretched this graph by a factor of five. 
			The two panels on the left use the same data, but the lower one accounts for asymmetric drift effects in the derivation of $\Lz$ by taking $R_g = \Lz/\meanVg = \Lz/ (226 \kms)$ instead of $R_g = \Lz/V_{c,\odot} = L_z/(238 \kms) $. The panels on the right again use the same data as the lower left panel for positive $z$ (upper right panel) and negative $z$ (lower right panel).}
		\label{fig:katz}
	\end{figure*}

	\subsection{Connecting to the vertical wave pattern}
	\label{sec:LzW}
	In Fig.~\ref{fig:katz} we show the $\meanUg$-$\Lz$ pattern overlaid with the same evaluation for $\meanWg$ vs. $\Lz$. 
	We further introduce the guiding centre radius $\Rg = \Lz/\Vc$, where for the sake of simplicity and easy reproduction we assume a flat rotation curve with $V_c =  238 \kms$. We duplicated our analysis using the more realistic rotation curve by \citep{McMillan17}. Not surprisingly, the resulting differences are negligible: Using a different $V_c = 233.1 \kms$ just shifts the pattern proportionally in $\Lz$; the change of $V_c$ with $R$ is much smaller than the width of the observed structures and smaller than the systematic uncertainties from the distance measurements. 
	
	We instantly see that the vertical wave-pattern detected by \cite{schonrich2018warp} appears to be strongly correlated with the $\meanUg$ wave discussed in this paper: the wave-peaks in $\meanWg$ appear to line up with the outer troughs of the $\meanUg$ wave. This could be a mere coincidence, but the degree of alignment suggests rather a physical correlation. As the amplitude in $\meanWg$ is a factor $\sim 5$, our sample could not yield a reliable signal for $\meanWg$ vs. $\phi$ to track the phase shift. 
	
	Many hypotheses can be drawn for this alignment. Of course a common origin could be suggested if both the radial and the vertical wave originate from the same dwarf galaxy impact \citep[e.g.][]{binney2018origin}. 
	This may be related to the phase space spiral being observed in both vertical and radial motion, but on the other hand, in a naive picture, it is not clear, why the alignment is then not removed by different frequencies in radial vs. vertical oscillation. The same complication would arise, if one wanted to adopt the less favoured explanation of the phase space spiral by a bar buckling event \citet{spiralbar2019}.
	\citet{monari2016Spiral} showed that the presence of spiral arms leads to vertical breathing modes aligned with the radial mean motion, which however include a vertical velocity field of odd parity that is not observed here. 
	Another predicition has been made in \citet{massettagger97}. In their paper, the authors predict that near an outer Lindblad resonance a spiral wave should interact with the galactic warp to produce a coupled horizontal and vertical wave. However, this effect should again be antisymmetric around the Galactic plane. 
	
	We thus examine the signals separated by galactic hemisphere in Fig.~\ref{fig:katz}. We find a mild difference of the vertical and radial velocity patterns above and below the Galactic plane, with a mildly stronger overlap of the extrema in $\meanUg$ and $\meanWg$ for positive $z$. While $\meanUg$ is slightly decreased in the Northern hemisphere (both when taken with respect to $R$ and to $\Lz$), $\meanWg$ is shifted upwards in our plot and shows a borderline significant third peak near $\Rg \sim 7.2 \kpc$. The general alignment is however clearly visible for both hemispheres.
	
	Fig.~\ref{fig:katz} also compares the $\meanUg$ vs. $R$ pattern to the $\meanUg$-$\Lz$ dependence.
	Trends of $\meanUg$ with $R$ have been predicted at least since the time of Hipparcos \citep[][]{Muehlbauer03}. Later, hints for a gradient in $\meanUg$ vs. $R$ have been found e.g. by \cite{Siebert2012Spiral} in RAVE data. They were explained by \cite{faure2014Spiral}, who showed with a 3D test-particle simulation that a spiral structure can be responsible for the observed pattern. \cite{Liu2018Spiral} continued along the line of argumentation by \cite{Muehlbauer03} and suggested that a bar is responsible for the dependence of $\meanUg$ on $R$. The simple pattern that we find in $\meanUg$ vs. $R$ very strongly resembles and qualitatively agrees with this prediction from \cite{Muehlbauer03}.
	We also note that the $\meanUg$ vs. $R$ dependence in Fig.~\ref{fig:katz} resembles the findings in fig. 12 of \citet{GaiaKatz}. It is also evident that this dependence is mostly a washed-out (by epicyclic motions of the orbits around their guiding centre) version of what we see in $\Lz$. To make this clearer, we repeat the plot of the top panel of Fig.~\ref{fig:katz} in the bottom panel, with one change: instead of using the assumed circular velocity $V_c = 238 \kms$ to translate $\Lz$ to $\Rg$, we add an asymmetric drift and use $\meanVg = 226 \kms$ to translate. Unsurprisingly, this results in a good agreement between the patterns in $\Lz$ and $R$. 
	A plot showing that the washing out of the pattern is also observed in radial against azimuthal velocity can be found in Fig.~\ref{fig:VgUgPlot} of the appendix. 
	
	\section{Conclusion} \label{sec:conclusion}
	We have for the first time closely examined a prominent wave-like pattern in mean radial velocities $\meanUg$ vs. angular momentum $\Lz$ in the RV subsample of Gaia DR2. The amplitude of the pattern is of order $4 \kms$ with a wavelength of about $285 \kms \kpc$ in $\Lz$ or $\sim 1.2 \kpc$ in guiding centre radius $\Rg$. To our knowledge, this pattern has not been directly predicted by any Galaxy model, though of course the offsets in $\meanUg$ predicted in the $\Vg$-$\meanUg$ plane by papers like \cite{Monari19} for distinct entities like the Hercules stream should be related to our finding.
	
	The structure is evidently not bound to a single Galactocentric radius $R$. We can detect it in a region of $\sim 7 - 10 \kpc$, i.e. a band of $\sim 3 \kpc$ around the Sun. There is a slight drift of the ridges towards smaller $\Lz$ for small $\Vg$, however the dependence of $\Lz$ on $\Vg$ and $R$ makes it hard to raise definite claims for the evolution of the pattern with these quantities. 
	
	The wave-like structure is not limited to kinematically cold thin-disc stars. We binned the sample in the local vertical energy $\Ez$ and found that the wave-like pattern persists to stars with $\Ez \sim 1200 (\kms)^2$, i.e. stars with orbits passing beyond thick disc altitudes $|z| > 1 \kpc$. For these kinematically very hot populations, the amplitude appears to be roughly half of the amplitude for kinematically cold stars, but this decrease is likely over-estimated due to larger distance uncertainties blurring the pattern. Towards larger $\Ez$ the pattern shifts to the left, i.e. smaller $\Lz$ with a trend of about $0.05 \kpc\kms / (\kms)^2$. This is likely connected to the slower radial frequencies of stars at higher altitudes.
	
	Most interestingly, the Gaia dataset is sufficiently extended that we can detect the slope of the ridges against the Galactic angle $\phi$. Without specifying a model, we can thus draw conclusions on the physical cause of the pattern: All ridges tend towards lower $\Lz$ for larger $\phi$, consistent with a trailing pattern in Galactic azimuth. However, there is a clear deviation with the peak in positive $\Ug$ near $\Lz \sim 2100 \kms \kpc$ showing only about half the change of the rest of the pattern. If we filter out this peak from our analysis, the slope of the remaining pattern in $\Lz$ vs. $\phi$ is consistent with an $m = 4$ wave-number.
	
	Hence, it is logical to associate this uniformly shifted part with an $m = 4$ excitation, possibly caused by the spiral pattern of the Milky Way or higher order ($m \geq 4$) contributions to the bar potential. The not (or only slightly) sloped broad peak hints to a connection with a bar resonance, however as to which depends on the bar pattern speed. A fast bar with a pattern speed of $ \sim 55 \kms\kpc^{-1}$ as for example advocated for in \citet{Dehnen00} and \citet{Antoja14Bar}  would have its outer Lindblad resonance in this region. However, a proposed much slower pattern speed of $39 \kms \kpc^{-^1}$ \citep[e.g.][]{Monari19, Perez17} connects it rather to the corotation resonance. In this area falls also the highly debated origin of the Hercules moving group, that has often been associated with the outer Linblad resonance \citep{Dehnen00}, whereas some recent studies suggest a corotation resonance origin \citep{monari2019hercules,donghia2019hercules}.
	There are nonetheless no properties of the pattern that favour one or the other bar model.

	Comparing the $\meanUg$ vs. $\Rg$ pattern to the similar pattern in $\meanUg$ vs. $R$, we find that the overall peak in $\meanUg$ vs. $R$ (which is already present in the analysis of \cite{GaiaKatz}) around $R \sim 8.8 \kpc$ is in line with the $\meanUg$ vs. $\Rg$ pattern being smeared out by epicyclic motions, once we account for the fact that the latter is shifted towards slightly larger values of $\Rg$ by the asymmetric drift.
	
	Last, we compared the newly found $\Ug$ vs. $\Lz$ pattern to the wave-like structure found by \cite{schonrich2018warp} in DR1. This pattern is still present in Gaia DR2, but its wavelength in $\Lz$ is longer than the wavelength of the radial velocity pattern. The two peaks in $W$ vs. $\Lz$ clearly visible in this selection fall roughly in the region of two troughs of $\Lz$. 
	There is a hint of a third peak in $\meanW$ coinciding with the lowest $\meanUg$ trough in $\Lz$, particularly visible in the northern galactic hemisphere. 
	
	\section{Acknowledgements}
	
	It is a pleasure to thank J. Binney, W. Dehnen, and R. Chiba for helpful comments. JF thanks the Stiftung Maximilianeum, the Ev. Studienwerk Villigst and the Max-Weber-Programm for their support
	and Merton College Oxford for their hospitality, without which this work would never have been undertaken. RS is supported by a Royal Society University Research Fellowship. This work was performed using the Cambridge Service for Data Driven Discovery (CSD3),
	part of which is operated by the University of Cambridge Research Computing on behalf of the STFC DiRAC HPC Facility (www.dirac.ac.uk). The DiRAC component of CSD3 was funded by BEIS capital funding via STFC capital grants ST/P002307/1 and ST/R002452/1 and STFC operations grant ST/R00689X/1. DiRAC is part of the National e-Infrastructure. This work has made use of data from the European Space Agency (ESA) mission {\it Gaia} (\url{https://www.cosmos.esa.int/gaia}), processed by the {\it Gaia} Data Processing and Analysis Consortium (DPAC, \url{https://www.cosmos.esa.int/web/gaia/dpac/consortium}). Funding for the DPAC has been provided by national institutions, in particular the institutions participating in the {\it Gaia} Multilateral Agreement.

	

	\bibliographystyle{mnras}
	\bibliography{main.bib}
	
	
	\appendix

	\section{Sample Selection and Quality cuts}
	
	\begin{figure}
		
		\centering
		
		\epsfig{file=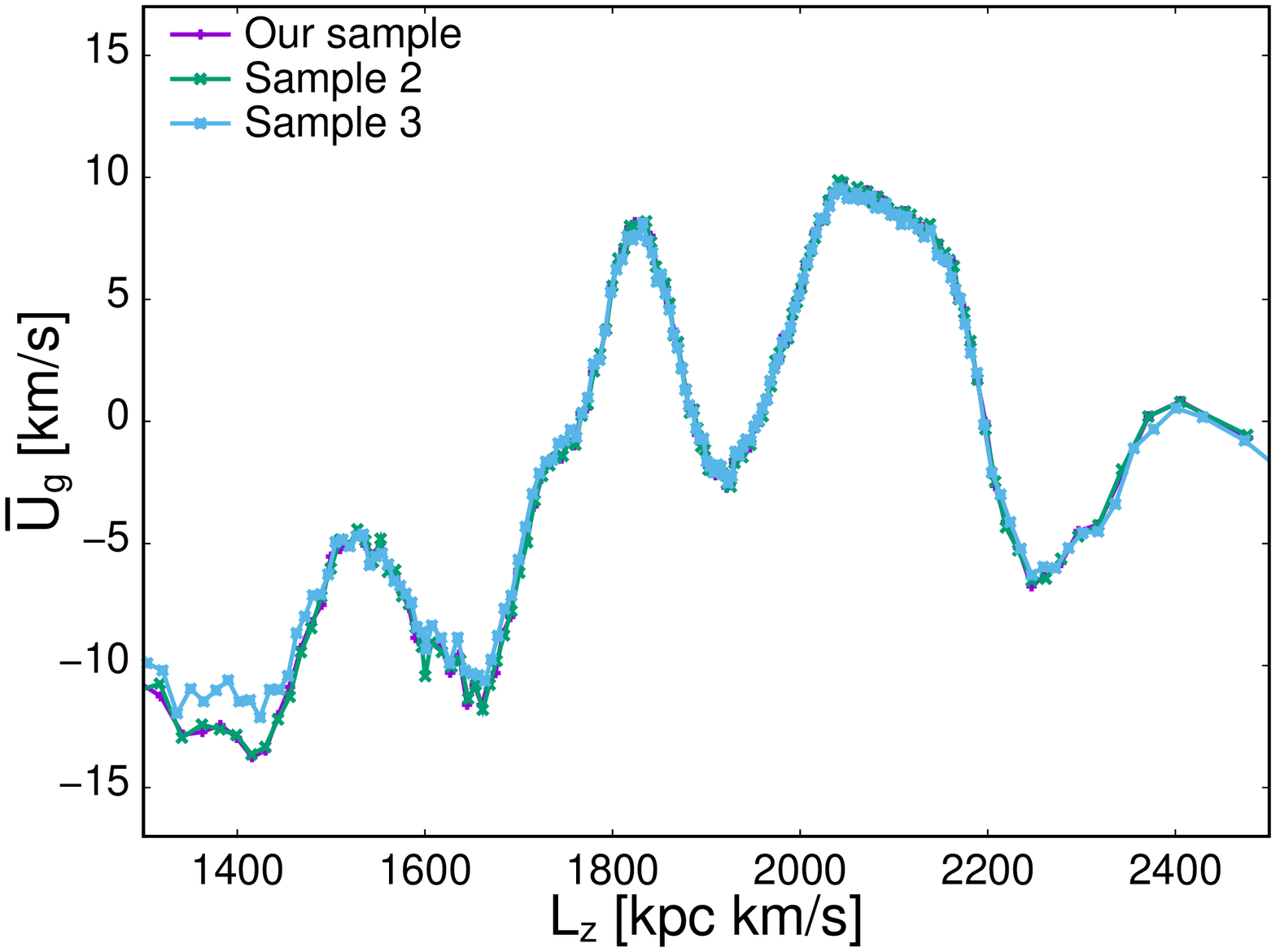,angle=-0,width=\hsize}
		
		\caption{Comparison of the $\Lz-\Ug$ pattern for three data samples derived from Gaia data.
			Sample 2 also increases the parallax error $\sigma_p = \sqrt{\sigma{'}_p^2 + (\delsigpar)^2}$ by $\delsigpar = 0.043 \mas$ but corrects for an parallax offset $\delta p = \sigma_p$.
			Sample 3 only includes a parallax offset $\delta p = 0.048 \mas $ and no increase of the parallax error, i.e. $\delsigpar = 0$. It is the most extended, but also the one with the largest relative uncertainties (the formal error uncertainty in that sample is up to a factor $2$ smaller for most stars), so the signal in $\Lz$ will be a little more blurred.
			Further information on the derivation of our samples can be found in \citet{S19b}}
		\label{fig:SampleComp}
	\end{figure}

	We dedicate this Appendix to some further quality tests in order to ascertain the significance of our findings and test for possible systematic effects.
	First, we want to make sure that the found pattern in $\Lz-\meanUg$ is not dependent on the choice of assumptions on parallax error $\delsigpar$ and parallax bias $\delpar$ in Gaia DR2. In the main text, we used exclusively the sample with $\delsigpar = 0.043 \mas $ and $\delpar = 0.054 \mas$. Fig.~\ref{fig:SampleComp} demonstrates that choosing any sample has only marginal consequences for the derived pattern of $\meanUg$ vs. $\Lz$, with a minor deviation for very small $\Lz$ in the third sample, which uses $\delsigpar = 0$ and $\delpar = 0.048 \mas$. This is easily understood, since the lack of an additional error term in $\sigpar$ allows for a further extent of the sample under our quality cut $p/\sigpar > 10$. We also note  that this sample has to be treated with caution, since there is good evidence that the Gaia pipeline values for $\sigpar$ are underestimated.

	\begin{figure}
		\centering
		\epsfig{file=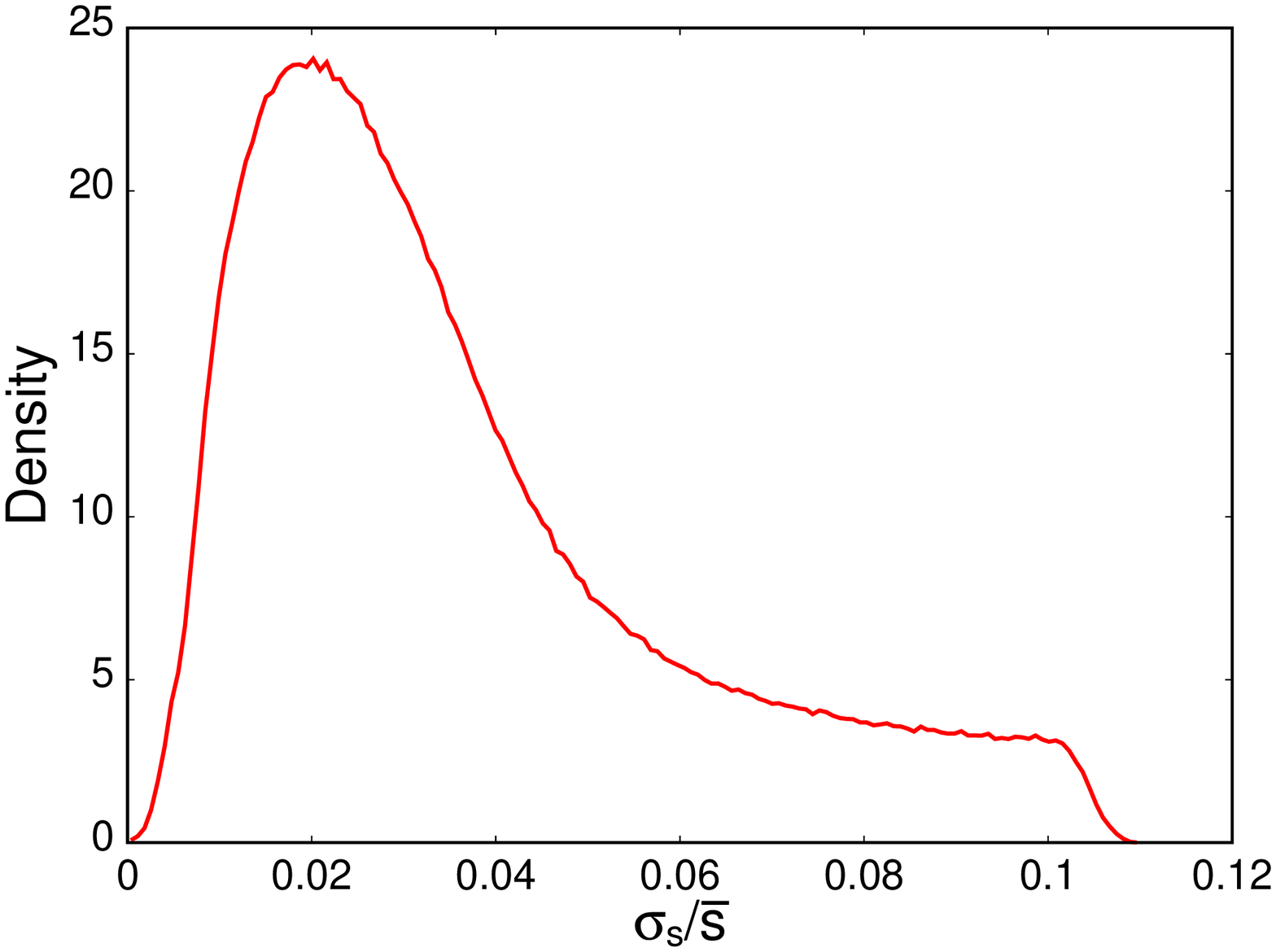,angle=-0,width=\hsize}
		\caption{Distribution of distance standard deviation over estimated distance for our sample after applying the quality cuts}
		\label{fig:GaiaQuality}
	\end{figure}

	Fig.~\ref{fig:GaiaQuality} shows the distribution of the standard deviation of the distances to estimated distance ratio, to help the reader assess the impact of distance uncertainties on our results. The relative error $\sigma_s / \overline{s}$ of the expected distance $\overline{s}$ peaks at around $2 \%$. The quality cut $p/ \sigpar > 10$ of course results in a similar cutoff in $\sigma_s / \overline{s}$, removing basically all stars with an estimated relative distance uncertainty great than $\sim 11 \%$. The strong peak at small relative errors results from the magnitude based selection in the Gaia RV subset, which strongly favours nearby stars. Note that the large distance uncertainties are of course found predominantly near the distant rim of our sample, the consequence of which we will examine further below.

	\begin{figure}
		\centering
		\epsfig{file=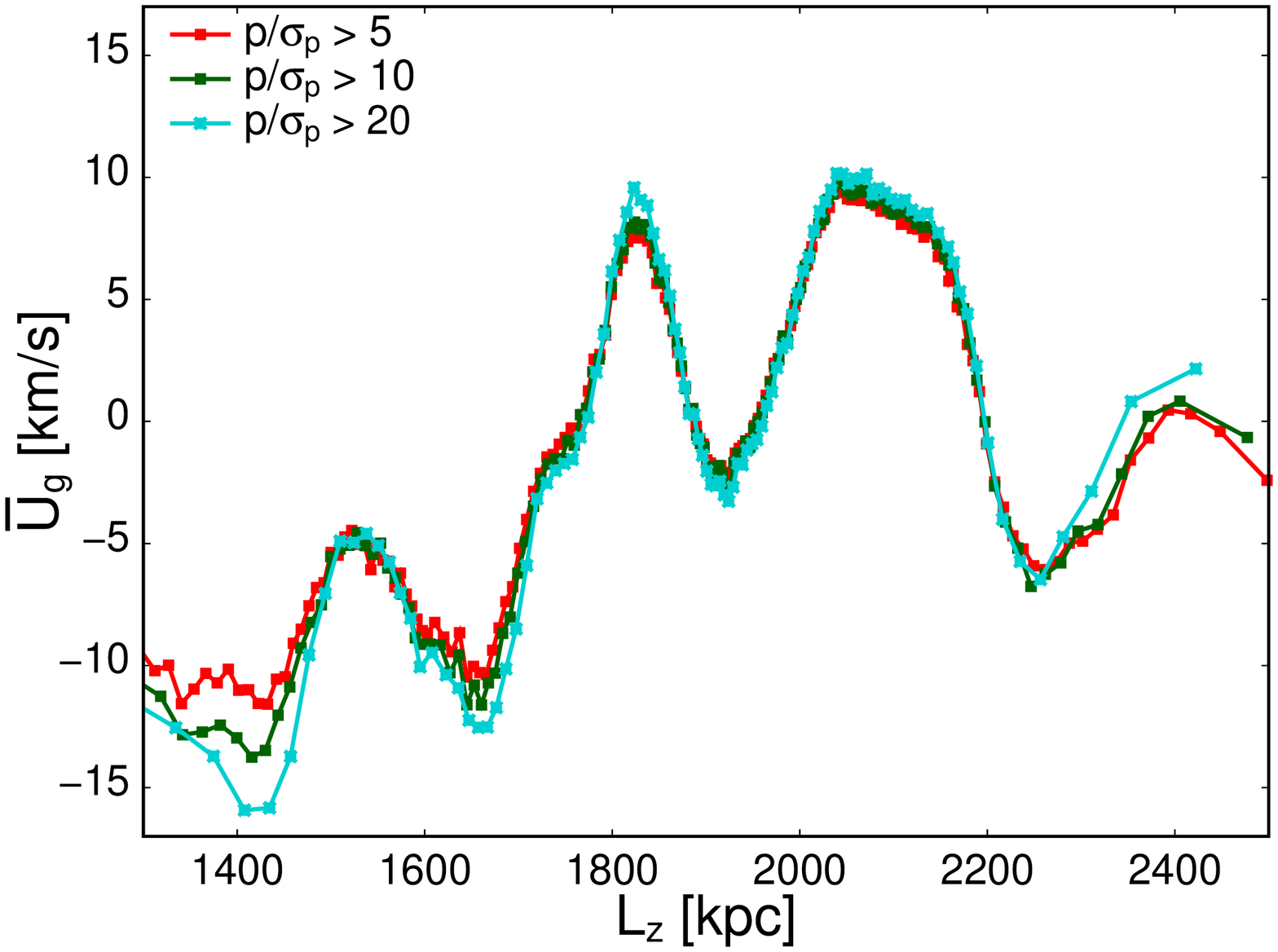,angle=-0,width=\hsize}
		\caption{Comparison of the $\Lz-\meanUg$ pattern for different quality cuts in parallax over parallaxerror.}
		\label{fig:ParallaxQuality}
	\end{figure}
	
	It is important to consider how different demands on the quality of our sample affect the pattern. Throughout this work, we required (in addition to the quality cuts suggested in section 8 of \citet{S19b}) a quality cut $p/ \sigpar > 10$. However, as Fig.~\ref{fig:ParallaxQuality} shows, different $p/\sigma_p$ ratios have only very little influence on the strength of the pattern, with a slightly lower amplitude for lower quality requirements due to random errors.
	
\begin{figure}
		\centering
		\epsfig{file=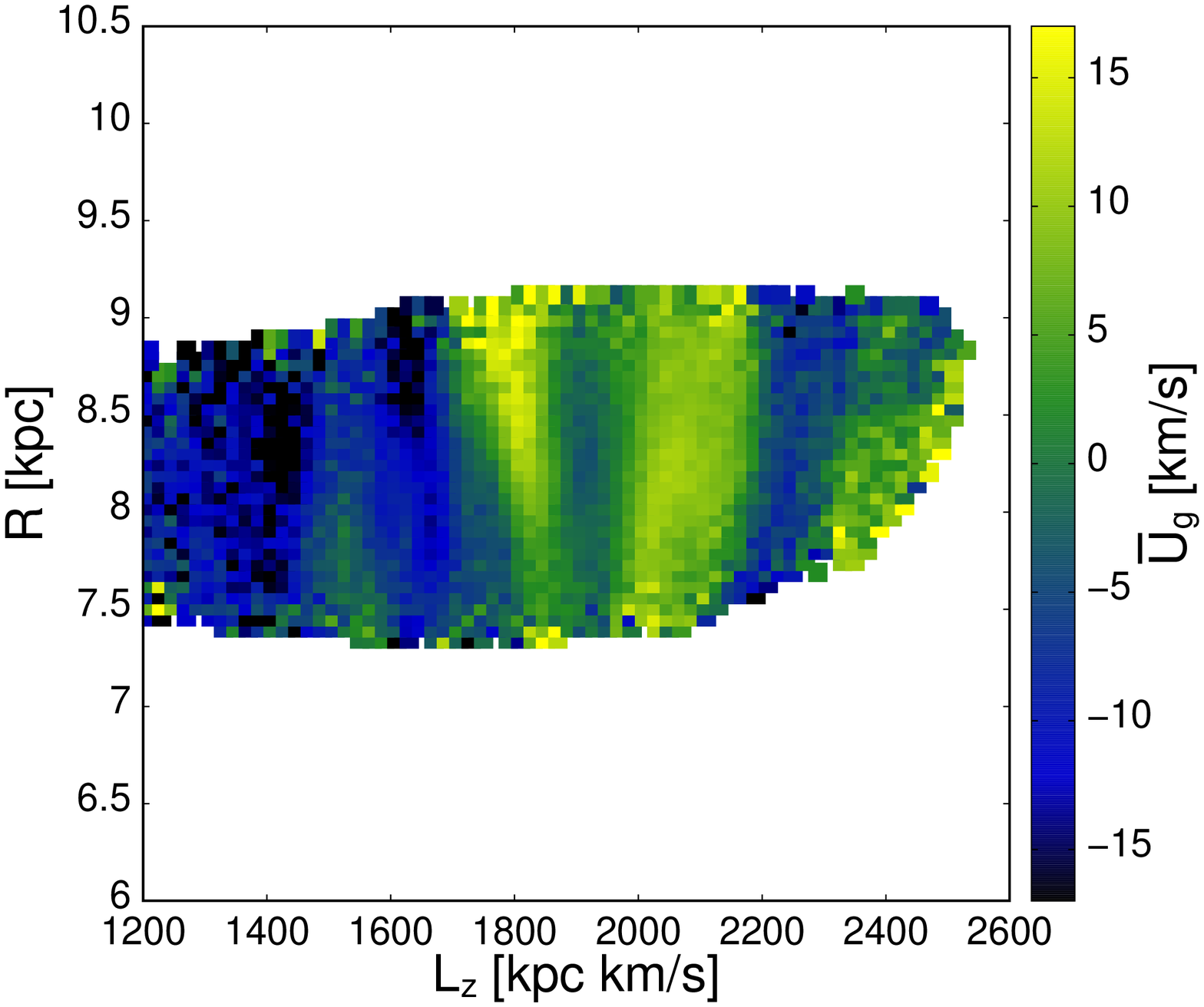,angle=-0,width=\hsize}
		\epsfig{file=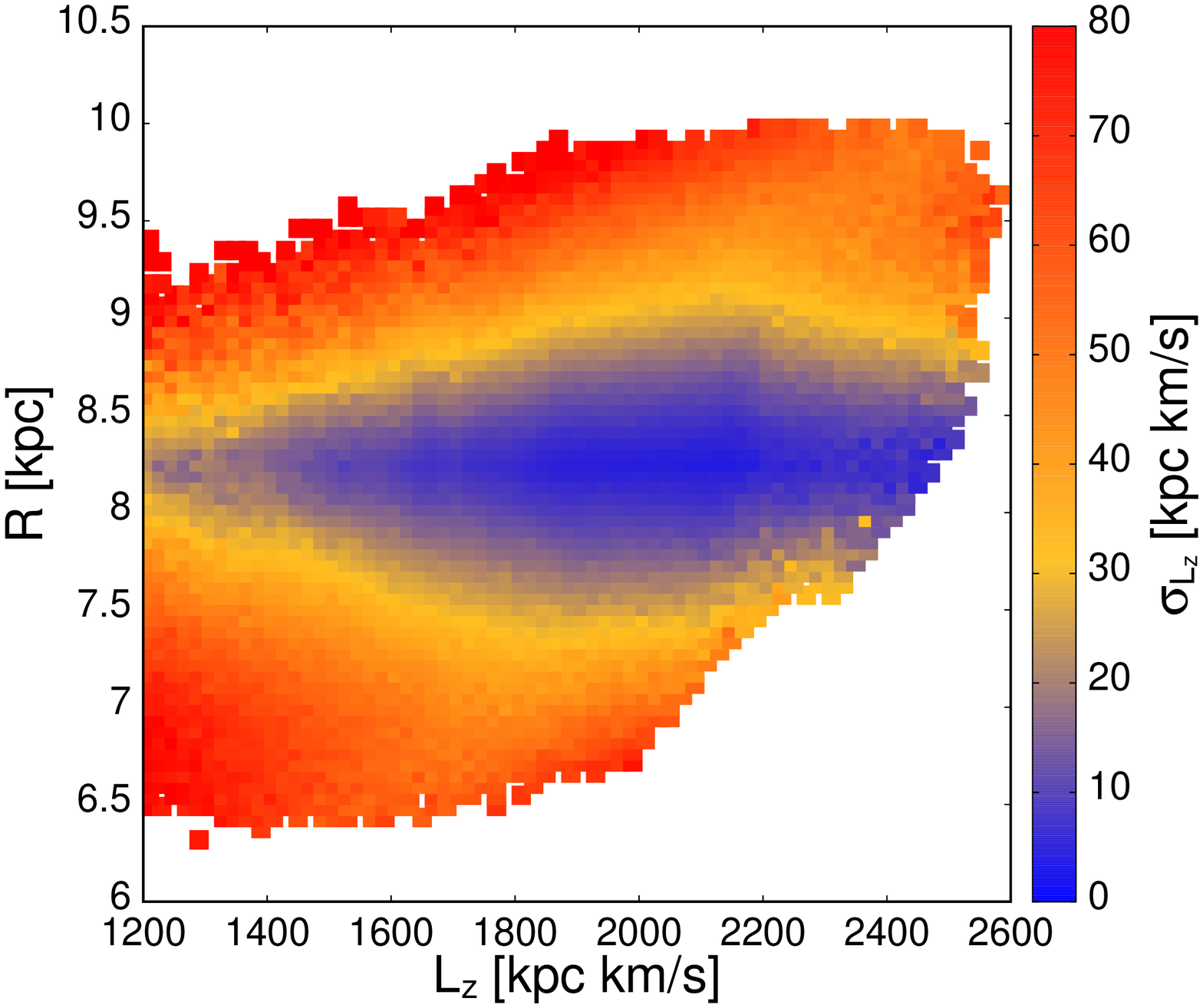,angle=-0,width=\hsize}
		\caption{Upper Panel: Same Plot as in the lower panel of Fig.~\ref{fig:LzEz} but with the higher quality requirement of parallax over parallax error bigger 20.
			Lower panel: Estimated standard deviation of the angular momentum $\Lz$ from the distance error.}
		\label{fig:LzRQuality}
	\end{figure}

	As our discussed wave pattern proved especially hard to resolve in radius, we repeat the plot of the bottom panel of Fig.~\ref{fig:LzEz} in the upper panel of Fig.~\ref{fig:LzRQuality} with a parallax cut of $p/\sigma_p > 20$. We find that the radial data range is strongly affected by this requirement and that the peculiar bent in $\meanUg$ for radii of $\sim 7.5 \kpc$ cannot be resolved anymore. It is thus likely that a definitive analysis of this has to await Gaia DR3. 
	
	As discussed in section \ref{sec:patternradialvsangular}, any pattern in the angular momentum should be increasingly washed out with increasing $\Lz$ uncertainty and should nearly disappear for
	an uncertainty exceeding roughly half a wavelength of the pattern. Stars at small $R$ are particularly prone due to the additive effect of a distance error both on $R$ and $\Vg$ in this region.
	We tried to estimate the error in $\Lz$ from the distance uncertainty by taking $\sigma_{\Lz} \approx \sigma_s \cdot d \Lz / d s $, where $s$ is the expected distance to the star and $\sigma_s$ the standard deviation of the distance probability distribution. The resulting pattern of the averaged $\sigma_{\Lz}$ (in quadrature) in the $\Lz$--$R$ plane can be found in the lower panel of Fig.~\ref{fig:LzRQuality}.
	Consistent with our findings in the main text, in the red regions of this figure we do not expect to still find a clear $\meanUg$-signal, as a $\sigma_{\Lz} > 70 \kpc \kms$ exceeds one quarter of the wave-length of the observed pattern. 
	
	\begin{figure}
		\centering
		\epsfig{file = 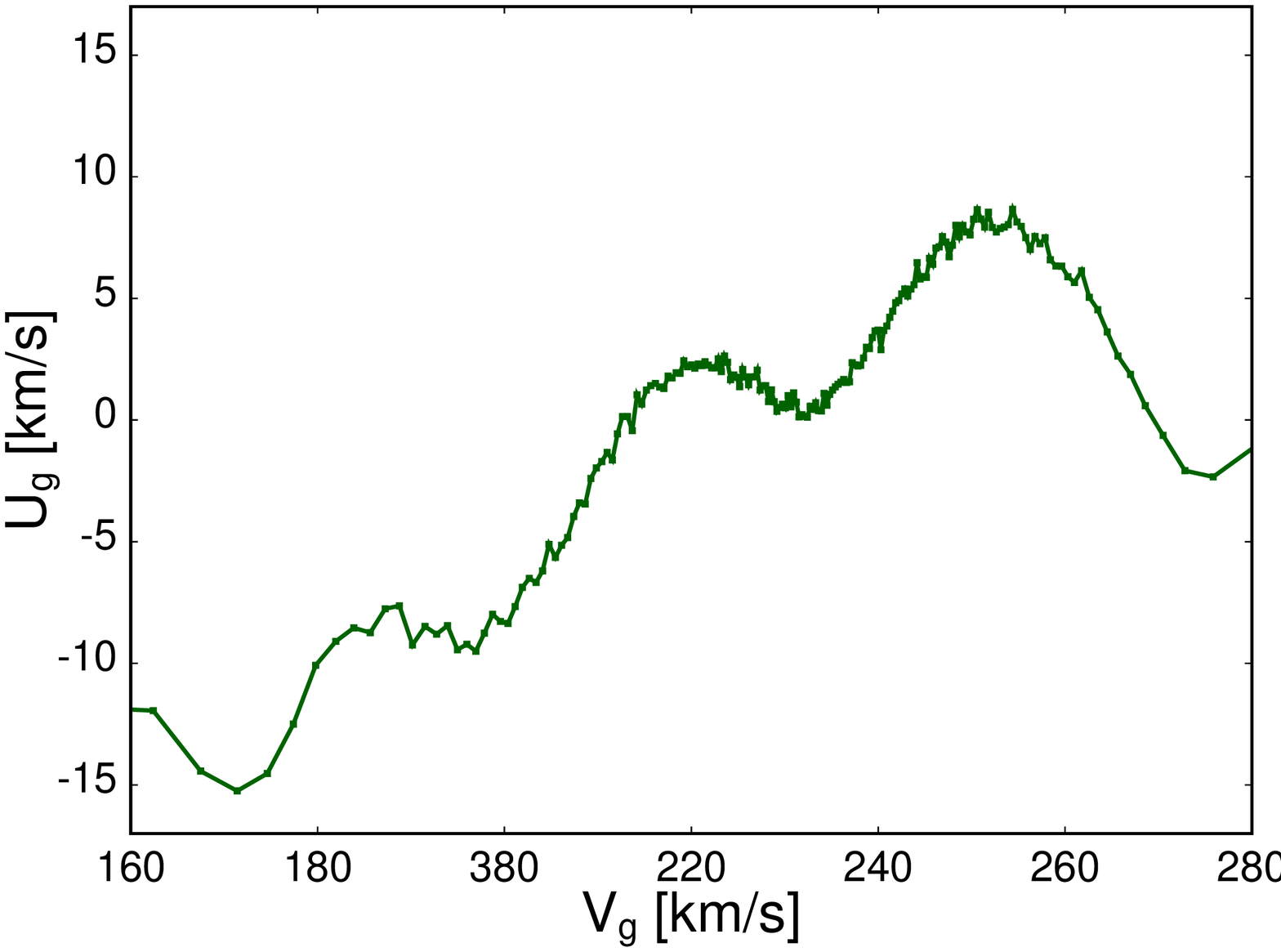, angle=-0, width= \hsize}
		\caption{Azimuthal vs. radial velocity for the sample with $p/\sigma_p >10$}
		\label{fig:VgUgPlot}
	\end{figure}	
	
	Following the discussion in \ref{sec:LzW} that a washed out version of the discussed wave pattern can be observed also in radial velocity versus radius (Fig.~\ref{fig:katz}), Fig.~\ref{fig:VgUgPlot} shows that the same effect takes place, if we consider $\Ug$ against the azimuthal velocity $\Vg$ instead of the angular momentum. The remaining pattern however follows the behaviour of $\Lz-R$ yet more closely.


	\bsp	
	\label{lastpage}
\end{document}